\documentclass[aps,prx,twocolumn,superscriptaddress,floatfix]{revtex4-2}

\usepackage{graphicx}
\usepackage{dcolumn}
\usepackage{bm}
\usepackage{amssymb}
\usepackage{amsmath}
\usepackage{microtype}
\usepackage{xfrac}
\usepackage{array}
\usepackage[dvipsnames]{xcolor}

\newcommand{\angstrom}{\text{\normalfont\AA}}

\begin{document}

\title{The case for a U(1)$_\pi$ Quantum Spin Liquid Ground State \\ in the Dipole-Octupole Pyrochlore Ce$_2$Zr$_2$O$_7$}

\author{E.~M.~Smith}
\affiliation{Department of Physics and Astronomy, McMaster University, Hamilton, ON, L8S 4M1, Canada}
\affiliation{Brockhouse Institute for Materials Research, McMaster University, Hamilton, ON L8S 4M1, Canada}

\author{O.~Benton}
\affiliation{Max Planck Institute for the Physics of Complex Systems, N\"{o}thnitzer Str. 38, Dresden 01187, Germany}

\author{D.~R.~Yahne}
\affiliation{Department of Physics, Colorado State University, 200 W. Lake St., Fort Collins, CO 80523-1875, USA}

\author{B.~Placke}
\affiliation{Max Planck Institute for the Physics of Complex Systems, N\"{o}thnitzer Str. 38, Dresden 01187, Germany}

\author{R.~Sch\"{a}fer}
\affiliation{Max Planck Institute for the Physics of Complex Systems, N\"{o}thnitzer Str. 38, Dresden 01187, Germany}

\author{J.~Gaudet}
\affiliation{Department of Physics and Astronomy, McMaster University, Hamilton, ON, L8S 4M1, Canada}

\author{J.~Dudemaine}
\affiliation{D\'epartement de Physique, Universit\'e de Montr\'eal, Montr\'eal, Canada}
\affiliation{Regroupement Qu\'eb\'ecois sur les Mat\'eriaux de Pointe (RQMP)}

\author{A.~Fitterman}
\affiliation{D\'epartement de Physique, Universit\'e de Montr\'eal, Montr\'eal, Canada}
\affiliation{Regroupement Qu\'eb\'ecois sur les Mat\'eriaux de Pointe (RQMP)}

\author{J.~Beare}
\affiliation{Department of Physics and Astronomy, McMaster University, Hamilton, ON, L8S 4M1, Canada}

\author{A.~R.~Wildes}
\affiliation{Institut Laue-Langevin, 71 Avenue des Martyrs CS 20156, 38042 Grenoble Cedex 9, France}

\author{S.~Bhattacharya}
\affiliation{Laboratoire de Physique des Solides, CNRS, Univ. Paris-Sud, Universit\'e Paris-Saclay, 91405 Orsay Cedex, France}

\author{T.~DeLazzer}
\affiliation{Department of Physics, Colorado State University, 200 W. Lake St., Fort Collins, CO 80523-1875, USA}

\author{C.~R.~C.~Buhariwalla}
\affiliation{Department of Physics and Astronomy, McMaster University, Hamilton, ON, L8S 4M1, Canada}

\author{N.~P.~Butch}
\affiliation{Center for Neutron Research, National Institute of Standards and Technology, MS 6100 Gaithersburg, Maryland 20899, USA}

\author{R.~Movshovich}
\affiliation{Los Alamos National Laboratory, Los Alamos, New Mexico 87545, USA}

\author{J.~D.~Garrett}
\affiliation{Brockhouse Institute for Materials Research, McMaster University, Hamilton, ON L8S 4M1, Canada}

\author{C.~A.~Marjerrison}
\affiliation{Brockhouse Institute for Materials Research, McMaster University, Hamilton, ON L8S 4M1, Canada}

\author{J.~P.~Clancy}
\affiliation{Department of Physics and Astronomy, McMaster University, Hamilton, ON, L8S 4M1, Canada}
\affiliation{Brockhouse Institute for Materials Research, McMaster University, Hamilton, ON L8S 4M1, Canada}

\author{E.~Kermarrec}
\affiliation{Laboratoire de Physique des Solides, CNRS, Univ. Paris-Sud, Universit\'e Paris-Saclay, 91405 Orsay Cedex, France}

\author{G.~M.~Luke}
\affiliation{Department of Physics and Astronomy, McMaster University, Hamilton, ON, L8S 4M1, Canada}
\affiliation{Brockhouse Institute for Materials Research, McMaster University, Hamilton, ON L8S 4M1, Canada}

\author{A.~D.~Bianchi}
\affiliation{D\'epartement de Physique, Universit\'e de Montr\'eal, Montr\'eal, Canada}
\affiliation{Regroupement Qu\'eb\'ecois sur les Mat\'eriaux de Pointe (RQMP)}

\author{K.~A.~Ross}
\affiliation{Department of Physics, Colorado State University, 200 W. Lake St., Fort Collins, CO 80523-1875, USA}
\affiliation{Canadian Institute for Advanced Research, 661 University Ave., Toronto, ON, M5G 1M1, Canada.}

\author{B.~D.~Gaulin}
\affiliation{Department of Physics and Astronomy, McMaster University, Hamilton, ON, L8S 4M1, Canada}
\affiliation{Brockhouse Institute for Materials Research, McMaster University, Hamilton, ON L8S 4M1, Canada}
\affiliation{Canadian Institute for Advanced Research, 661 University Ave., Toronto, ON, M5G 1M1, Canada.}

\date{\today}

\begin{abstract} 
The Ce$^{3+}$ pseudospin-$\frac{1}{2}$ degrees of freedom in the pyrochlore magnet Ce$_2$Zr$_2$O$_7$ are known to possess dipole-octupole (DO) character, making it a candidate for novel quantum spin liquid (QSL) ground states at low temperatures. We report new polarized neutron diffraction at low temperatures, as well as heat capacity ($C_p$) measurements on single crystal Ce$_2$Zr$_2$O$_7$. The former bears both similarities and differences from that measured in the canonical dipolar spin ice compound Ho$_2$Ti$_2$O$_7$, while the latter rises sharply at low temperatures, initially plateauing near 0.08~K, before falling off towards a high temperature zero beyond 3~K. Above $\sim$0.5~K, the $C_p$ data set can be fit to the results of a quantum numerical linked cluster (NLC) calculation, carried out to 4$^{\mathrm{th}}$ order, that allows estimates for the terms in the near-neighbour XYZ Hamiltonian expected for such DO pyrochlore systems. Fits of the same theory to the temperature dependence of the magnetic susceptibility and unpolarized neutron scattering complement this analysis. A comparison between the resulting best fit NLC calculation and the polarized neutron diffraction shows both agreement and discrepancies, mostly in the form of zone-boundary diffuse scattering in the non-spin flip channel, which are attributed to interactions beyond near-neighbours. The lack of an observed thermodynamic anomaly and the constraints on the near-neighbour XYZ Hamiltonian suggest that Ce$_2$Zr$_2$O$_7$ realizes a U(1)$_\pi$ QSL state at low temperatures, and one that likely resides near the boundary between dipolar and octupolar character. 

\end{abstract}
\maketitle

\section{Introduction}

The rare-earth pyrochlore oxides, R$_2$B$_2$O$_7$, where $R^{3+}$ is a trivalent rare-earth ion and $B^{4+}$ is a non-magnetic tetravalent transition-metal ion, display a wealth of both exotic and conventional magnetic ground states. Their $R^{3+}$ ions decorate a network of corner-sharing tetrahedra, one of the archetypes for geometrical frustration in three dimensions, and this crystalline architecture underlies many of their exotic properties~\cite{GardnerReview2010}. A separation of energy scales, with crystal electric field (CEF) effects dominating over exchange interactions, often results in a well-separated CEF ground state doublet for the $R^{3+}$ ion, and interacting pseudospin-$\frac{1}{2}$ degrees of freedom at low temperatures~\cite{GingrasReview2014, Hallas2018, RauReview2019}.

It is well-appreciated that the CEF Hamiltonian determines both the size of the magnetic moment at the $R^{3+}$ site and its anisotropy, but less well-appreciated that the symmetry of the CEF ground-state can imprint itself on the exchange Hamiltonian~\cite{RauReview2019,Li2017,Huang2014}. The possible symmetries of the ground state doublets then lead to an important classification of the rare earth pyrochlores, which depends on how their CEF doublet transforms under time-reversal symmetry and the point group symmetry of the $R^{3+}$ site. Three classes of doublets arise, one for non-Kramers ions with an even number of electrons, and two for Kramers ions with an odd number of electrons. The non-Kramers case gives rise to a pseudospin wherein one component of the pseudospin transforms as a magnetic dipole and two transform as quadrupoles. For Kramers ions, we have the familiar case where all three components of the pseudospin in the ground state doublet transform as magnetic dipoles, as well as the more exotic one where two components transform as magnetic dipoles and one transforms as an octupole. This latter case is known to describe the CEF Kramer's ground state of 4\textit{f}$^1$ Ce$^{3+}$ in Ce$_2$Zr$_2$O$_7$~\cite{Gaudet2019,Gao2019}, a dipole-octupole (DO) ground state doublet, and also that of its sister pyrochlore, Ce$_2$Sn$_2$O$_7$~\cite{Sibille2020}. Fig.~1(a) pictorially displays the magnetic charge distributions associated with both magnetic dipoles and octupoles decorating the tetrahedra on part of a cubic pyrochlore lattice. As discussed above, for the dipole-octupole doublets relevant to Ce$_2$Zr$_2$O$_7$, a single component of the pseudospin-$\frac{1}{2}$ degree of freedom (the $y$-component) behaves as an octupole, while the $x$ and $z$ components behave as dipoles under the symmetry of the lattice and time-reversal symmetry, as schematically illustrated in Fig.~1(b).

Such DO doublets decorating pyrochlore lattices are theoretically known to allow for at least 6 distinct quantum disordered and ordered ground states, with three in each of the dipole and octupole sectors~\cite{Benton2020,Patri2020, Huang2018b}. Recent neutron scattering measurements on single crystal Ce$_2$Zr$_2$O$_7$ have uncovered a signal that strongly resembles predictions for the energy-integration of emergent photon excitations in a U(1) quantum spin ice~\cite{Gaudet2019}, while recent experiments on powder samples of Ce$_2$Sn$_2$O$_7$ have been interpreted in terms of a U(1) quantum spin ice ground state in the octupole sector~\cite{Sibille2020}. 

\begin{figure}[t]
\linespread{1}
\par
\includegraphics[width=3.4in]{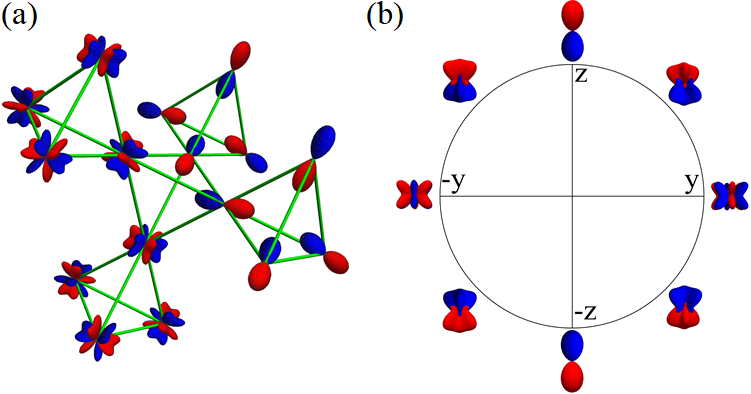}
\par
\caption{(a) The magnetic charge distributions associated with octupoles (left) and dipoles (right) are depicted at the vertices of five corner-sharing tetrahedra, making up part of the pyrochlore lattice. (b) Octupolar and dipolar components inhabit the same pseudospin-$\frac{1}{2}$ Ce$^{3+}$ degrees of freedom in Ce$_2$Zr$_2$O$_7$, such that $y$-components behave as octupoles, while the $x$ and $z$ components of each pseudospin-$\frac{1}{2}$ behave as dipoles, as schematically illustrated here using the magnetic charge distributions associated with different directions of pseudospin in the $yz$-plane.} 
\label{Figure1}
\end{figure}

\section{Outline of the paper}

In this paper, we present new polarized neutron diffraction and heat capacity measurements on single crystal Ce$_2$Zr$_2$O$_7$. The former bears both similarities and differences from that measured in the canonical dipolar spin ice compound, Ho$_2$Ti$_2$O$_7$, while the latter shows no sign of a thermodynamic phase transition above $T$ = 0.06~K. $C_p$ rises sharply at low temperatures, initially plateauing near 0.08~K, before falling off towards a high temperature zero beyond 3~K, consistent with previous measurements~\cite{Gao2019}. We have modelled the high temperature $C_p$, and the powder-averaged magnetic susceptibility using quantum numerical linked cluster (NLC) expansions. This allows us to estimate and constrain the parameters of the anticipated near-neighbour XYZ Hamiltonian. To the extent that interactions beyond near-neighbour do not alter ground state selection, we constrain the nature of the ground state itself, with the results indicating a U(1)$_{\pi}$ QSL ground state is selected at low temperature. 

We use the resulting near neighbour exchange parameters to calculate the equal-time spin flip (SF) and non-spin flip (NSF) structure factors in the $[HHL]$ scattering plane. This calculation resembles the new polarized neutron diffraction measurements in the SF channel from single crystal Ce$_2$Zr$_2$O$_7$, but cannot account for the observed zone-boundary diffuse scattering in the NSF channel. We attribute this discrepancy to interactions beyond near-neighbour in the Hamiltonian, which are expected to be small, and a full study of which is beyond the scope of our present work. The same discrepancy exists for spin-polarized neutron diffraction from Ho$_2$Ti$_2$O$_7$, where it was ascribed to expected long range dipolar interactions \cite{Fennell2009}. NLC calculations using the same near-neighbour exchange Hamiltonian were also carried out to 7$^{\mathrm{th}}$ order. While these agree with the 4$^{\mathrm{th}}$ order calculations above $\sim$0.5 K, they depart from the measured $C_p$ at lower temperatures. We interpret this as arising from the same interactions beyond near neighbour in Ce$_2$Zr$_2$O$_7$ that were revealed by the NSF zone boundary scattering. As these are relatively weak, they only manifest themselves at low temperatures.

A further consistency check is carried out via semiclassical Monte Carlo and Molecular Spin Dynamics using the best fit near-neighbour Hamiltonian. This calculation accounts for the energy dependence of the inelastic spectral weight making up the diffuse scattering at low temperatures without further adjustment of the NLC-determined near-neighbour Hamiltonian. We further show that the full $R\ln(2)$ entropy of the DO ground state doublet can be accounted for to 10~K with a smooth extrapolation of $C_p$ from the lowest temperature data point at $T$ = 0.06~K, to zero at $T$ = 0~K, using a theoretical form which is simultaneously consistent with both the expected behavior of a U(1) QSL at low temperature, and the high temperature limit of the NLC calculations. Interestingly, the Pauling, classical spin ice entropy $R\ln(2)$ less $\frac{R}{2}\ln(\frac{3}{2}$) is recovered from the peak in the $C_p$ data at $\sim$0.08 K, to 10 K.

\section{Polarized Neutron Diffraction}

We have carried out new polarized diffraction measurements on single crystal Ce$_2$Zr$_2$O$_7$ using the D7 diffractometer at the Institute Laue Langevin. This diffractometer employs a spin polarized monochromatic incident beam, which was $E_i$ = 3.47~meV for this experiment. This configuration effectively integrates over $\Delta E$ $\sim$0.16~meV during the course of a diffraction measurement. A single polarization direction, perpendicular to the $[HHL]$ scattering plane, was employed, and as such the spin flip (SF) and non-spin flip (NSF) diffuse scattering profiles can be independently measured. The diffuse scattering associated with these two cross sections, SF and NSF, are shown in the $[HHL]$ scattering plane for Ce$_2$Zr$_2$O$_7$ in Fig.~2(a) and 2(b), respectively for the temperature-difference data set $T$ = 0.045~K - $T$ = 10~K. For comparison, the corresponding SF and NSF diffuse scattering patterns as measured on single crystal Ho$_2$Ti$_2$O$_7$ at $T$ = 1.7~K are shown in Fig.~2(c) and 2(d), respectively~\cite{Fennell2009}. These earlier spin polarized diffuse scattering measurements on Ho$_2$Ti$_2$O$_7$ (Ref.~\cite{Fennell2009}) played a formative role in the development of classical spin ice physics, as they drew clear attention to ``pinch point" scattering within the SF cross section at (0,0,2) and (1,1,1) and equivalent wavevectors, due to the presence of a classical Coulomb phase at low temperature. These measurements on Ho$_2$Ti$_2$O$_7$ also observed zone-boundary diffuse scattering in the NSF channel, which was later attributed to the long range dipolar interactions relevant to the large Ho$^{3+}$ dipole moments.

\begin{figure}[]
\linespread{1}
\par
\includegraphics[width=3.4in]{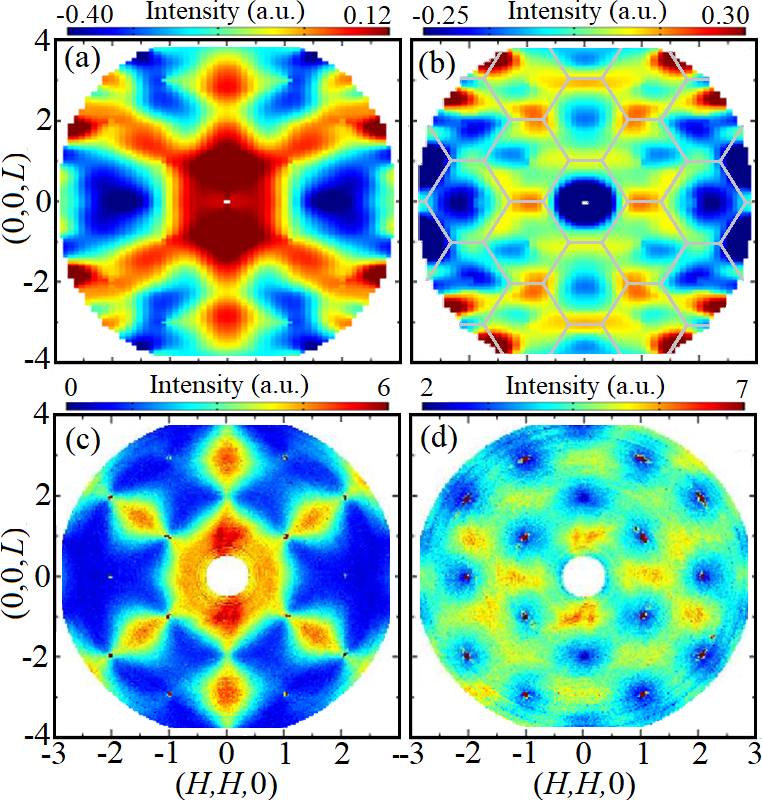}
\par
\caption{ The symmetrized $T$ = 45~mK - $T$ = 10~K temperature-difference neutron signal measured in the (a) SF and (b) NSF channels of our polarized neutron diffraction experiment on Ce$_2$Zr$_2$O$_7$. The (c) SF and (d) NSF scattering signals in the $[HHL]$ plane measured in a polarized neutron scattering experiment on Ho$_2$Ti$_2$O$_7$ at $T$ = 1.7~K \cite{Fennell2009}. The data in this figure is shown in arbitrary units.}
\label{Figure2}
\end{figure}

The comparison between the spin polarized diffuse scattering from Ce$_2$Zr$_2$O$_7$ and Ho$_2$Ti$_2$O$_7$ in Fig.~2 is interesting both in what is similar and where the discrepancies between the two materials lie. One may note however, that the comparison is made at quite different temperatures, 0.045 K for Ce$_2$Zr$_2$O$_7$ but only 1.7 K for Ho$_2$Ti$_2$O$_7$. In fact, the large Ho$^{3+}$ moments and effective ferromagnetic coupling cause Ho$_2$Ti$_2$O$_7$ to depolarize the beam at lower temperatures, whereas no such issue is present for Ce$_2$Zr$_2$O$_7$ due to its much smaller Ce$^{3+}$ moments. Quasi-pinch point SF scattering is observed near (0,0,2) Bragg positions in Ce$_2$Zr$_2$O$_7$, but it is not as constricted as that observed at (0,0,2) in Ho$_2$Ti$_2$O$_7$, even though the earlier measurements in Ho$_2$Ti$_2$O$_7$ were taken at much higher temperature. Furthermore while diffuse SF scattering extends out in (1,1,1) and equivalent directions in a snow flake-like pattern in Ce$_2$Zr$_2$O$_7$, pinch points appear to be absent in these directions.

In contrast, and somewhat surprisingly, the observed NSF diffuse scattering in Ce$_2$Zr$_2$O$_7$ is quite similar to that seen in Ho$_2$Ti$_2$O$_7$. In both cases the diffuse scattering tends to follow the face-centred cubic Brillouin zone boundaries, outlined in grey in Fig.~2(b). In Ho$_2$Ti$_2$O$_7$, this was ascribed to interactions beyond near neighbour~\cite{Fennell2009}, which was not surprising, given that dipolar interactions are expected to dominate over exchange interactions even for near neighbours in Ho$_2$Ti$_2$O$_7$. However, the Ce$^{3+}$ moments are $\sim$8 times smaller than those of Ho$^{3+}$ and hence dipolar interactions are expected to be $\sim$64 times smaller in Ce$_2$Zr$_2$O$_7$. We revisit our new polarized neutron diffraction data in Section V where we compare the measured SF and NSF signals to NLC calculations using the near-neighbour exchange parameters yielded in this work.

\section{Estimating the Near-Neighbour Exchange Parameters in the Spin Hamiltonian}

The gold standard for determining the microscopic spin Hamiltonian of magnetic materials is inelastic neutron scattering studies of spin wave spectra. This technique can and has been successfully applied to pyrochlore magnets with pseudospin-$\frac{1}{2}$ degrees of freedom arising from well-separated ground state CEF doublets, including Yb$_2$Ti$_2$O$_7$ and Er$_2$Ti$_2$O$_7$~\cite{Ross2011, Savary2012, Ross2014, Scheie2020, Thompson2017, Robert2015}. For disordered ground states, it is necessary to perform measurements in a sufficiently strong magnetic field, so as to polarize the ground state, thus giving rise to well defined spin wave spectra. However, this is not always possible. For example, the classical spin ice ground state as appears in Ho$_2$Ti$_2$O$_7$, doesn't allow transverse spin fluctuations - hence no well defined spin wave excitations are observed due to Ho$^{3+}$'s non-Kramers CEF doublet eigenvectors~\cite{Clancy2009}. No evidence for well defined spin waves has been observed to date in either zero or non-zero magnetic field in Ce$_2$Zr$_2$O$_7$, a likely consequence of the form of Ce$^{3+}$'s DO CEF ground state doublet and spin Hamiltonian. Hence estimates for the microscopic spin Hamiltonian parameters for such materials can only come from sophisticated modelling of other data, such as the high temperature thermodynamic data presented here.  We note that a related work has appeared coincident with this paper which performs independent modeling of heat capacity, magnetization and neutron scattering measurements in Ce$_2$Zr$_2$O$_7$, and reaches similar conclusions \cite{Changlani2021}.

\subsection{Introduction to the Exchange Parameters in the XYZ Hamiltonian}

The near-neighbour XYZ Hamiltonian appropriate to DO pyrochlores in a magnetic field may be written as~\cite{Li2017,Huang2014}, 

\begin{equation}
\begin{split}
    \mathcal{H}_\mathrm{XYZ} & = \sum_{<ij>}[     J_{\tilde{x}}{S_i}^{\tilde{x}}{S_j}^{\tilde{x}} + J_{\tilde{y}}{S_i}^{\tilde{y}}{S_j}^{\tilde{y}} + J_{\tilde{z}}{S_i}^{\tilde{z}}{S_j}^{\tilde{z}}] \\ 
    & - g_z \mu_\mathrm{B} \sum_{i} \mathbf{h} \cdot\hat{{\bf z}}_i({S_i}^{\tilde{z}}\cos\theta + {S_i}^{\tilde{x}}\sin\theta) \;\;.
\end{split}
\end{equation}

In this equation, ${S_{i}}^{\tilde{\alpha}}$ ($\alpha = \tilde{x}$, $\tilde{y}$, $\tilde{z}$) are the pseudospin components of atom $i$ in the local $\tilde{x}$, $\tilde{y}$, $\tilde{z}$ coordinate frame. This coordinate frame arises from rotation of the local $x$, $y$, $z$ coordinate frame, with the $z$ anisotropy axis connecting near-neighbor tetrahedra in the pyrochlore structure, by $\theta$ about the $y$-axis~\cite{Li2017,Huang2014}. The magnetic field is denoted as $\mathbf{h}$, and $\hat{{\bf z}}_i$ is the local anisotropy axis for the site $i$. The g-factor $g_z$ is fixed by the wave functions of the lowest CEF doublet, giving $g_z = 2.57$ for Ce$^{3+}$~\cite{Gaudet2019, Gao2019, Sibille2020}. $S_i^{\tilde{x}}$ and $S_i^{\tilde{z}}$ are distinguished from $S_i^{\tilde{y}}$ by how they transform under the point group of the lattice and time-reversal symmetry. $S_i^{\tilde{x}}$ and $S_i^{\tilde{z}}$ transform like a magnetic dipole while $S_i^{\tilde{y}}$ transforms like a component of the magnetic octupole tensor, as schematically illustrated in Fig. 1.  

The nearest-neighbour exchange Hamiltonian in Eq.~(1) has only three independent exchange parameters $(J_{\tilde{x}},  J_{\tilde{y}},  J_{\tilde{z}})$ in zero magnetic field. Theory has predicted the ground-state phase diagram for such a zero-field XYZ Hamiltonian, uncovering both quantum spin liquid (QSL) as well as ordered ground states~\cite{Benton2020,Patri2020}. Each of these can have either dipolar or octupolar nature - a QSL phase has octupolar nature if $|J_{\tilde{y}}| > |J_{\tilde{x}}|,|J_{\tilde{z}}|$ and dipolar nature if $|J_{\tilde{z}}| > |J_{\tilde{y}}|$ or $|J_{\tilde{x}}| > |J_{\tilde{y}}|$. An ordered phase has octupolar nature if $J_{\tilde{y}} < J_{\tilde{x}},J_{\tilde{z}}$ and dipolar nature if $J_{\tilde{z}} < J_{\tilde{y}}$ or $J_{\tilde{x}} < J_{\tilde{y}}$. One final classification comes about for U(1) QSL ground-states, based on whether the U(1) flux that penetrates the hexagonal plaquettes embedded in the pyrochlore structure, is equal to 0 or $\pi$. This leads to a distinction between U(1)$_0$ and U(1)$_{\pi}$ QSLs. The aforementioned theoretical studies then uncover six phases within the ground-state phase diagram: all-in all-out (AIAO) order, U(1)$_0$ QSL, and U(1)$_{\pi}$ QSL, each of which can have dipolar or octupolar nature. A separate theory study has provided evidence for a small portion of the ground-state phase diagram corresponding to a $\mathbb{Z}_2$ QSL phase~\cite{Huang2018b}. It is worth noting that inter-Ce$^{3+}$ interactions beyond near-neighbour are allowed, but are expected to be weak. Long-range, three-dimensional dipolar interactions must be present in Ce$_2$Zr$_2$O$_7$, however, they are expected to be weak due to the small dipole moment associated with the Ce$^{3+}$ CEF ground state doublet in Ce$_2$Zr$_2$O$_7$ \cite{Gao2019,Gaudet2019}.  Exchange interactions beyond near neighbour are also expected to be weak due to the localized nature of 4\textit{f} electron wavefunctions in rare earth insulators.

\subsection{Heat Capacity and Numerical Linked Cluster Calculations}

\begin{figure}[t]
\linespread{1}
\par
\includegraphics[width=3.4in]{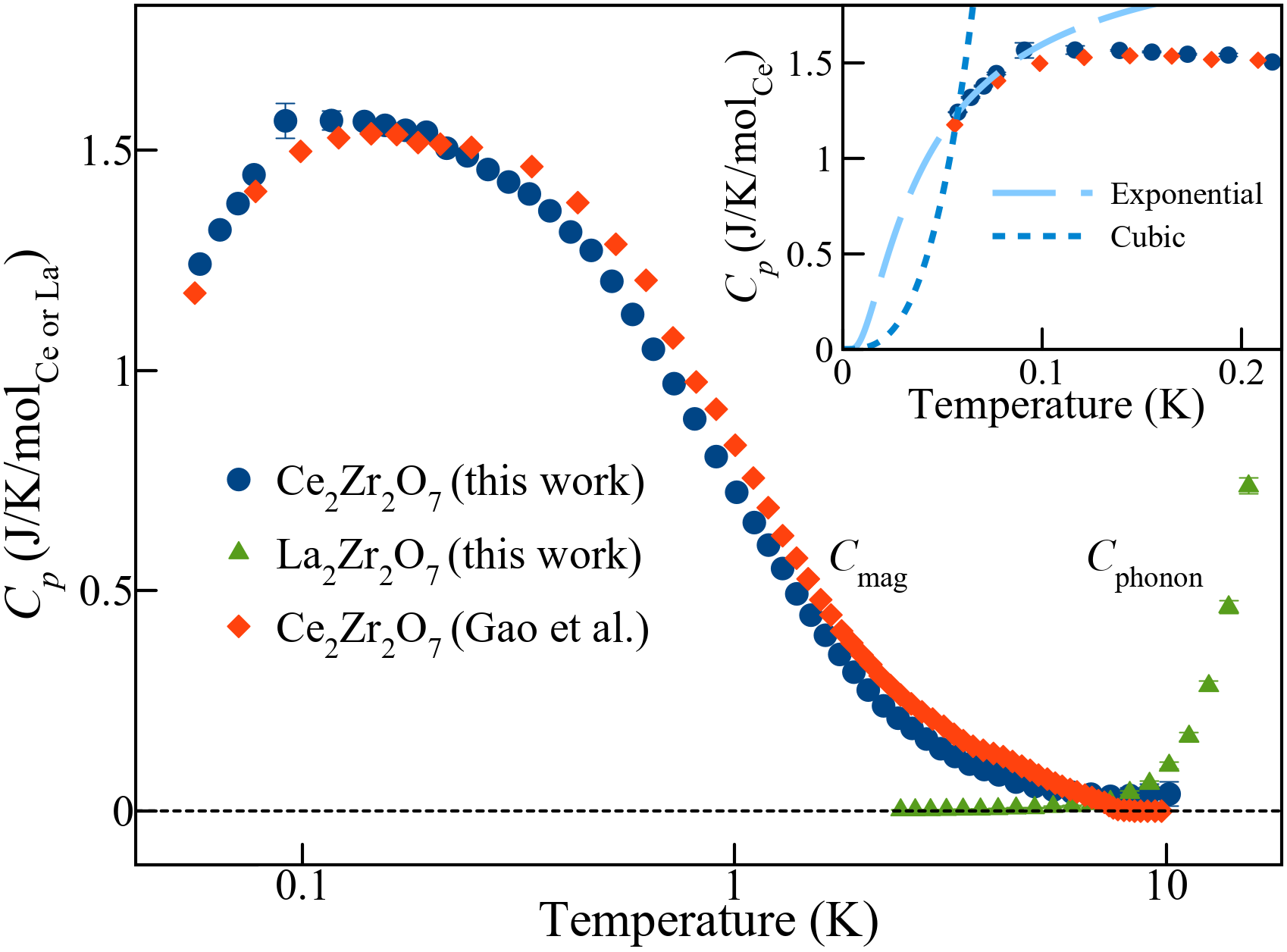}
\par
\caption{The magnetic contribution to the heat capacity ($C_\mathrm{mag}$) for the Ce$_2$Zr$_2$O$_7$ single crystal measured in the present work (blue) and in previous work by Gao \emph{et al}. (red)~\cite{Gao2019}. The phonon contribution to the heat capacity, estimated from measurements on a La$_2$Zr$_2$O$_7$ sample (green), was removed from $C_p$ to obtain $C_\mathrm{mag}$. The inset shows the best-fit simple exponential and cubic extrapolations to $T$ = 0~K for the present Ce$_2$Zr$_2$O$_7$ $C_\mathrm{mag}$. An exponential extrapolation, with an energy gap of $\sim$0.035~K, can smoothly connect to the finite temperature data, while a cubic extrapolation cannot.} 
\label{Figure3}
\end{figure}

The single crystal and powder samples of Ce$_2$Zr$_2$O$_7$ used in this study are from the same growth and synthesis employed in Ref.~\cite{Gaudet2019}. As reported there, stabilizing the Ce$^{3+}$ oxidation state in Ce$_2$Zr$_2$O$_7$ requires growth and annealing in strong reducing conditions to minimize the Ce$^{4+}$ content. The amount of sample oxidation (the value of $\delta$ in Ce$^{3+}_{2-2\delta}$Ce$^{4+}_{2\delta}$Zr$_{2}$O$_{7+\delta}$) can be tracked through x-ray diﬀraction measurements of the lattice parameter~\cite{Otobe2005}, and we estimate an oxidation level of $\delta$ $\sim$0.05 for the single crystal samples in the present work. Heat capacity measurements on a polished single crystal were carried out on a Quantum Design PPMS with dilution insert using the conventional quasi-adiabatic thermal relaxation technique. 

Heat capacity measurements were performed on our single crystal Ce$_2$Zr$_2$O$_7$ sample, along with a polycrystalline sample of La$_2$Zr$_2$O$_7$ (see Appendix A), which is used as a 4\textit{f}$^0$ analogue of Ce$_2$Zr$_2$O$_7$. The results are shown in Fig.~3, where the temperature axis is logarithmic. $C_p$ results on another Ce$_2$Zr$_2$O$_7$ single crystal from Ref.~\cite{Gao2019} are also overlaid for ease of comparison.  One can see that the phonon contribution to $C_p$, as measured in the La$_2$Zr$_2$O$_7$ sample, is negligible below $\sim$10~K, and thus $C_\mathrm{mag}$ is easily isolated. These results show that $C_\mathrm{mag}$ rises on decreasing temperature below $\sim$3~K, and then drops off sharply below $\sim$0.08~K, consistent with the earlier measurements (Ref.~\cite{Gao2019}) and a disordered ground state, as no sharp features associated with a phase transition can be identified. 

The order of the quantum NLC calculations, which were used to model the experimental results, refers to the maximum number of tetrahedra considered in a cluster. We have carried out NLC calculations for orders of 7 and less to model the magnetic heat capacity at temperatures above an order-dependent threshold. This threshold is set by the temperature above which the $n^{\mathrm{th}}$-order calculation for a particular set of near neighbour exchange parameters is consistent with the corresponding ($n$-1)$^{\mathrm{th}}$-order calculation. NLC calculations become progressively more time-consuming to carry out at higher order. For this reason, calculations of the high temperature $C_\mathrm{mag}$ with varying exchange parameters were carried out only to order 4, while calculations of other observables (integrated $S({\bf Q}, T)$ and susceptibility) were calculated at lower order. NLC calculations at order 7, the highest order reported here, were carried out for $C_\mathrm{mag}$ with a single set of exchange couplings only. Going beyond 6$^{\mathrm{th}}$ order is significant, because this is the first order at which the expansion contains non-trivial loops.

At temperatures of $T$ $\sim$0.5~K and above, the measured $C_\mathrm{mag}$ data can be compared with 4$^{\mathrm{th}}$ order NLC (NLC-4) calculations for $C_\mathrm{mag}$ in order to model and constrain Ce$_2$Zr$_2$O$_7$'s microscopic near-neighbour Hamiltonian. As the zero-field heat capacity contains no directional information, we define a new set of axes, $\{a,b,c\}$, to be the permutation of $\{\tilde{x}$, $\tilde{y}$, $\tilde{z}\}$ such that $|J_a|\geq|J_b|,|J_c|$ and $J_b \geq J_c$. This allows for a unique fit to $C_\mathrm{mag}$ but does not specify which values correspond to which exchange constants. Accordingly, the fit does not distinguish between the octupolar or dipolar nature of the ground-state. Nonetheless, knowledge of $J_{a}$, $J_{b}$, and $J_{c}$ suffices to determine whether the ground state is an ordered phase or a QSL phase~\cite{Benton2020}. 

This $J_{a}$, $J_{b}$, $J_{c}$ Hamiltonian can also be written in terms of raising and lowering operators with respect to ${S_i}^{a}$, giving:

\begin{equation}
\begin{split}
    \mathcal{H}_\mathrm{ABC} & = \sum_{<ij>}[J_{a}{S_i}^{a}{S_j}^{a} + J_{b}{S_i}^{b}{S_j}^{b} + J_{c}{S_i}^{c}{S_j}^{c}] \\
    & = \sum_{<ij>}[J_{a}{S_i}^{a}{S_j}^{a} - J_{\pm}({S_i}^{+}{S_j}^{-} + {S_i}^{-}{S_j}^{+}) \\
    & + J_{\pm\pm}({S_i}^{+}{S_j}^{+} + {S_i}^{-}{S_j}^{-})] \\
\end{split}
\end{equation}

\noindent in zero field, where $J_\pm = -\frac{1}{4}(J_b + J_c)$, $J_{\pm\pm} = \frac{1}{4}(J_b - J_c)$. 

\begin{figure}[t]
\linespread{1}
\par
\includegraphics[width=3.4in]{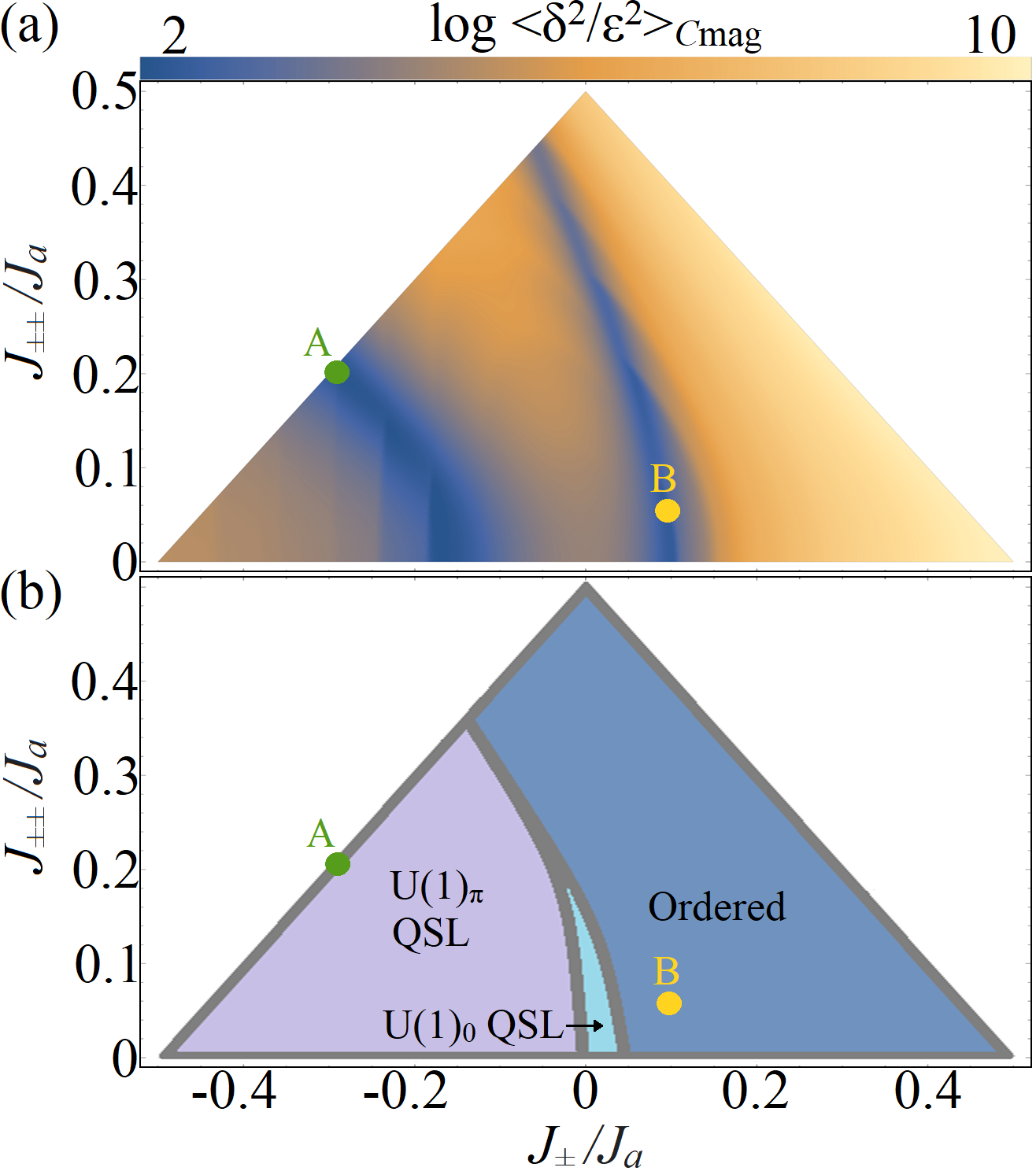}
\par
\caption{(a) The goodness-of-fit parameter ($\langle \frac{\delta^2}{\epsilon^2} \rangle_{C\mathrm{mag}}$) for the 4$^{\mathrm{th}}$-order NLC calculation compared to the measured $C_\mathrm{mag}$, as a function of the exchange parameters, $J_a$, $J_\pm$ = -$\frac{1}{4}$($J_b + J_c$), and $J_{\pm\pm}$ = $\frac{1}{4}$($J_b$ - $J_c$). This displays two local minima of $\langle \frac{\delta^2}{\epsilon^2} \rangle_{C\mathrm{mag}}$. The best-fit parameters are labelled as parameter set A and parameter set B. The global minimum corresponds to set A while set B is only locally optimal. (b) The best fit parameters from the NLC calculations (A and B) overlaid on the zero-field ground state phase diagram predicted for the XYZ model Hamiltonian and DO pyrochlores~\cite{Benton2020}. The set A exchange parameters are well-within the region of the phase diagram that is attributed to the U(1)$_{\pi}$ QSL, while the set B parameters are well-within the region attributed to an ordered ground state.} 
\label{Figure4}
\end{figure}

\begin{figure}[t]
\linespread{1}
\par
\includegraphics[width=3.4in]{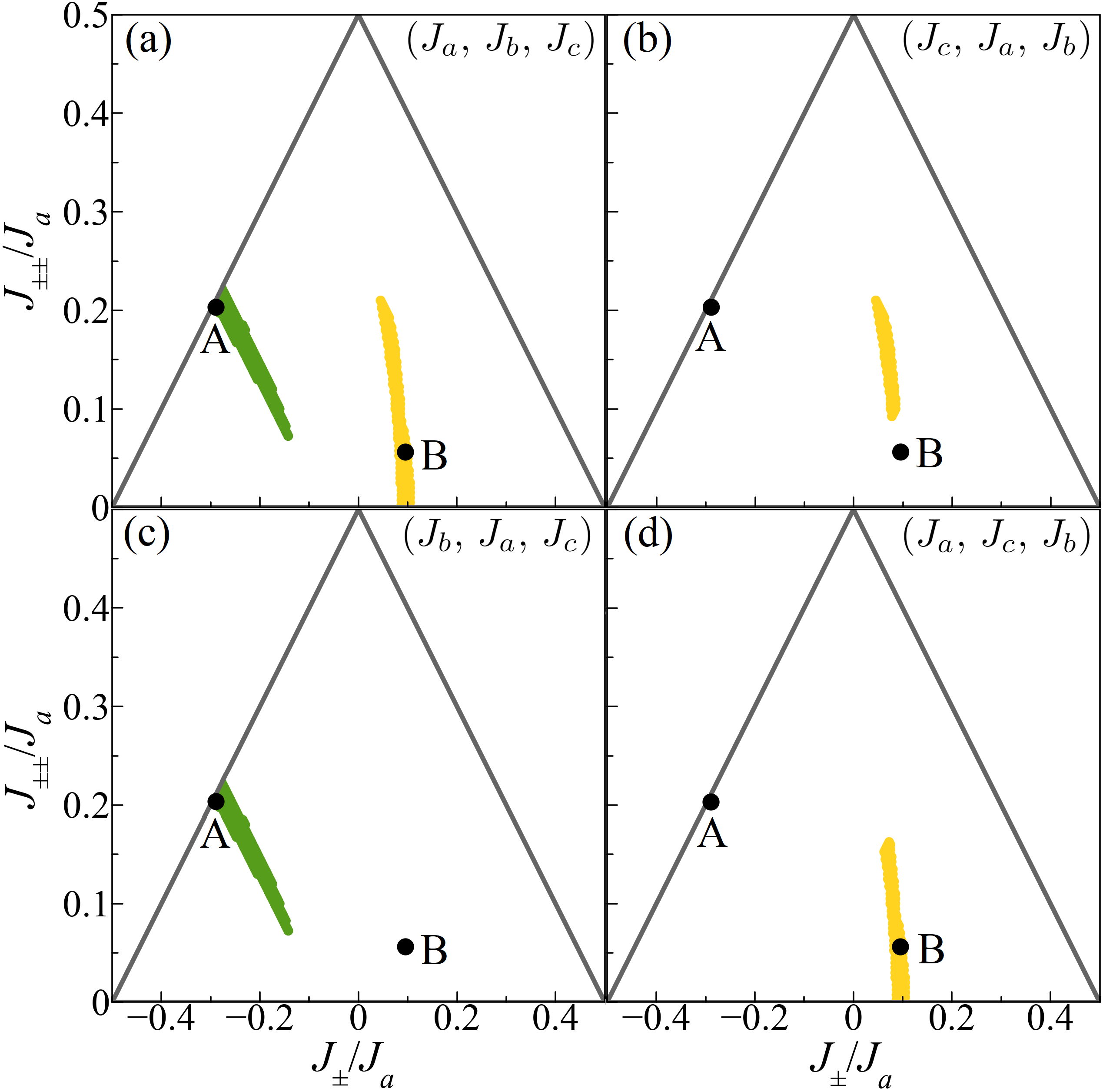}
\par
\caption{The regions of the XYZ phase diagram for which it is possible to obtain simultaneous reasonable NLC descriptions of $\chi$ and $C_\mathrm{mag}$ are indicated in green and yellow for $(J_{\tilde{x}}, J_{\tilde{y}}, J_{\tilde{z}})$ equal to the different permutations of $(J_a, J_b, J_c)$. We define the thresholds for reasonable $\chi$ and $C_\mathrm{mag}$ descriptions in Appendix C. (a) The regions of simultaneous $\chi$ and $C_\mathrm{mag}$ descriptions for the permutation in which $(J_{\tilde{x}}, J_{\tilde{y}}, J_{\tilde{z}})$ is equal to (a) $(J_a, J_b, J_c)$, (b) $(J_c, J_a, J_b)$, (c) $(J_b, J_a, J_c)$, and (d) $(J_a, J_c, J_b)$. The overall best-fit A parameters require that $(J_{\tilde{x}}, J_{\tilde{y}}, J_{\tilde{z}})$ is equal to $(J_a, J_b, J_c)$ or $(J_b, J_a, J_c)$; that is $J_{\tilde{z}} = J_c$.}
\label{Figure5}
\end{figure}

The set of exchange parameters $(J_a,  J_b,  J_c)$ best reproducing $C_\mathrm{mag}$ was obtained from a 4$^{\mathrm{th}}$ order NLC calculation with an Euler transformation to improve convergence. Heat capacity curves were calculated for values of -1 $\leq$ $J_b$ $\leq$ 1 and -1 $\leq$ $J_c$ $\leq$ $J_b$ in increments of 0.01, with $J_a$ = 1. Each curve was then re-scaled for best agreement with experiment to determine the value of $J_a$, according to the goodness-of-fit measure $\langle \frac{\delta^2}{\epsilon^2} \rangle_{C\mathrm{mag}} \propto \sum\frac{(C_\mathrm{mag}^\mathrm{NLC}(T_\mathrm{exp})-C_\mathrm{mag}^\mathrm{exp}(T_\mathrm{exp}))^2}{\epsilon(T_\mathrm{exp})^2}$; where the sum is over measured temperatures $T_\mathrm{exp}$ above the low-temperature threshold $\frac{0.7J_a}{k_\mathrm{B}}$, restricting the fit to the regime where the NLC calculations converge, and $\epsilon(T_\mathrm{exp})$ is the experimental uncertainty on the heat capacity at temperature $T_{exp}$. The values of $\langle \frac{\delta^2}{\epsilon^2} \rangle_{C\mathrm{mag}}$ over the entire phase space, after optimization of the scale $J_a$ for each parameter set, are shown in Fig.~4(a). This displays two extended regions in which there is good agreement with the experimental $C_\mathrm{mag}$. Both regions are entirely within one single phase in the predicted ground state phase diagram for the near neighbour XYZ model Hamiltonian (Fig.~4(b))~\cite{Benton2020}.

Some parameter sets within these regions can however be excluded due to their inability to describe the experimental magnetic susceptibility data. This is shown in Fig.~5 and explained in further detail in Subsection C of this section. The best fits within each region which are also consistent with the susceptibility data are found at the points $(J_a,  J_b,  J_c)$ = (0.064, 0.063, 0.011)~meV and (0.089, -0.007, -0.027)~meV, which we label as A and B, respectively. In Fig.~4(b) we overplot the optimal exchange parameters on top of the predicted ground state phase diagram for the near neighbour XYZ model Hamiltonian~\cite{Benton2020}. The set A (B) exchange parameters reside within the region corresponding to the $\pi$-flux U(1) QSL (ordered phase). Of these two parameter sets, parameter set A gives a better fit to the heat capacity. The calculated $C_\mathrm{mag}$'s using the 4$^{\mathrm{th}}$ order NLC with sets A and B are shown in Fig.~6. 

The 4$^{\mathrm{th}}$ order NLC $C_\mathrm{mag}$-calculation and fit was redone assuming 5$\%$ vacancies and $\langle \frac{\delta^2}{\epsilon^2} \rangle_{C\mathrm{mag}}$ again shows two locally optimal regions of parameter space. The best-fitting parameter sets that are also able to describe the measured susceptibility, A' and B', are very near to A and B in parameter space, respectively (see Appendix B). The global (local) minima at A' (B') lies within the region corresponding to the $\pi$-flux U(1) QSL (ordered phase). We therefore conclude that these results are robust to the presence of at least 5$\%$ Ce$^{4+}$ in Ce$_2$Zr$_2$O$_7$.

7$^{\mathrm{th}}$-order NLC (NLC-7) calculations for $C_\mathrm{mag}$ converge above $\sim$0.2 K, and these have been carried out for the optimal, set A, near-neighbour exchange parameters, as shown in the inset to Fig.~6. These higher order calculations are consistent with the NLC-4 calculations above $\sim$0.5 K. However at temperatures between $\sim$0.2 K and $\sim$0.5 K, the NLC-7 calculations do not quantitatively describe the measured $C_\mathrm{mag}$. We attribute this to interactions not included in the XYZ Hamiltonian (Eqs. (1) and (2)), those beyond near-neighbour, which are relatively weak and therefore only manifest themselves at the lowest temperatures. This is also consistent with the zone-boundary diffuse scattering observed in the NSF structure factor discussed above and shown in Fig.~2. Including the next-nearest-neighbour (NNN) part of the dipole-dipole interaction in the NLC-7 calculation did not significantly improve the agreement between theory and experiment, suggesting that either dipole-dipole interactions beyond NNN or additional exchange interactions are important.

\begin{figure}[t]
\linespread{1}
\par
\includegraphics[width=3.4in]{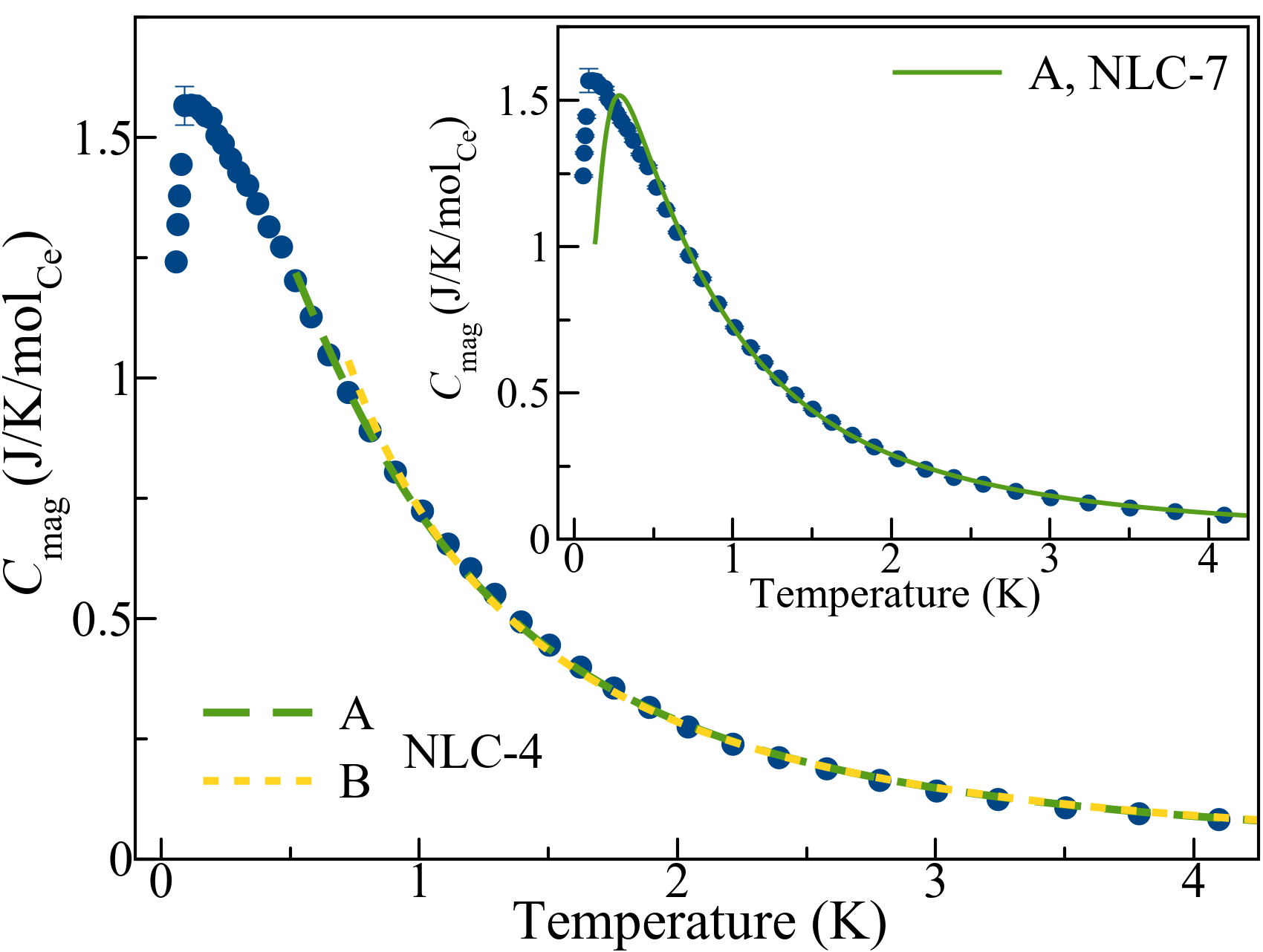}
\par
\caption{The results of the 4$^{\mathrm{th}}$-order NLC $C_\mathrm{mag}$-calculation for zero sample-oxidation, using the near-neighbour exchange parameters $J_a$ = 0.064~meV, $J_b$ = 0.063~meV, $J_c$ = 0.011~meV (set A) and $J_a$ = 0.089~meV, $J_b$ = -0.007~meV, $J_c$ = -0.027~meV (set B), overlaid on top of the measured $C_\mathrm{mag}$ for our Ce$_2$Zr$_2$O$_7$ sample. The inset shows the results of the 7$^{\mathrm{th}}$ order NLC $C_\mathrm{mag}$-calculation for zero sample-oxidation, using the set A near-neighbour exchange parameters.}
\label{Figure6} 
\end{figure}

\subsection{DC Magnetic Susceptibility}

While the zero-field $C_\mathrm{mag}$ contains no directional information, the temperature-dependent DC magnetic susceptibility ($\chi$) does because it is sensitive to the magnetic moment, which distinguishes between pseudospin components. Specifically, $\chi$ is dependent on the values of $J_{\tilde{x}}$, $J_{\tilde{y}}$, $J_{\tilde{z}}$, and $\theta$. A 2$^{\mathrm{nd}}$ order NLC expansion (NLC-2) is used to calculate $\chi$ (see Appendix C). Specifically, we use NLC-2 to fit measurements of $\chi$ from a powder sample of Ce$_2$Zr$_2$O$_7$ in order to narrow down the possible parameter sets and to distinguish between possible permutations of the exchange parameters. 

As mentioned above, some parameter sets within the region of good agreement for $C_\mathrm{mag}$ cannot be made to agree with $\chi$, for any choice of $\theta$ or permutation of parameters, and are therefore excluded. Fig.~5 shows the regions of the phase diagram for which it is possible to obtain simultaneous agreement with $C_\mathrm{mag}$ and $\chi$, for $(J_{\tilde{x}},  J_{\tilde{y}},  J_{\tilde{z}})$ equal to the different permutations of $(J_a,  J_b,  J_c)$. 

For the B parameters, we can rule out the possibility of $J_{\tilde{z}}$ being the largest exchange parameter, and we find different optimal values of $\theta$ for the remaining permutations. For the A parameter set, and all nearby parameter sets for which a good fit can be found, the results of the NLC-2 fitting to $\chi$ suggest that $\theta \sim 0$ and that $J_{\tilde{z}}$ is the weakest exchange parameter as Fig.~5 and Fig.~7 demonstrate. Accordingly, the only allowed permutations of exchange parameters from the A set satisfy $J_{\tilde{x}} \sim J_{\tilde{y}}$, implying that Ce$_2$Zr$_2$O$_7$ resides near the boundary between dipolar and octupolar nature. 

\begin{figure}
\linespread{1}
\par
\includegraphics[width=3.4in]{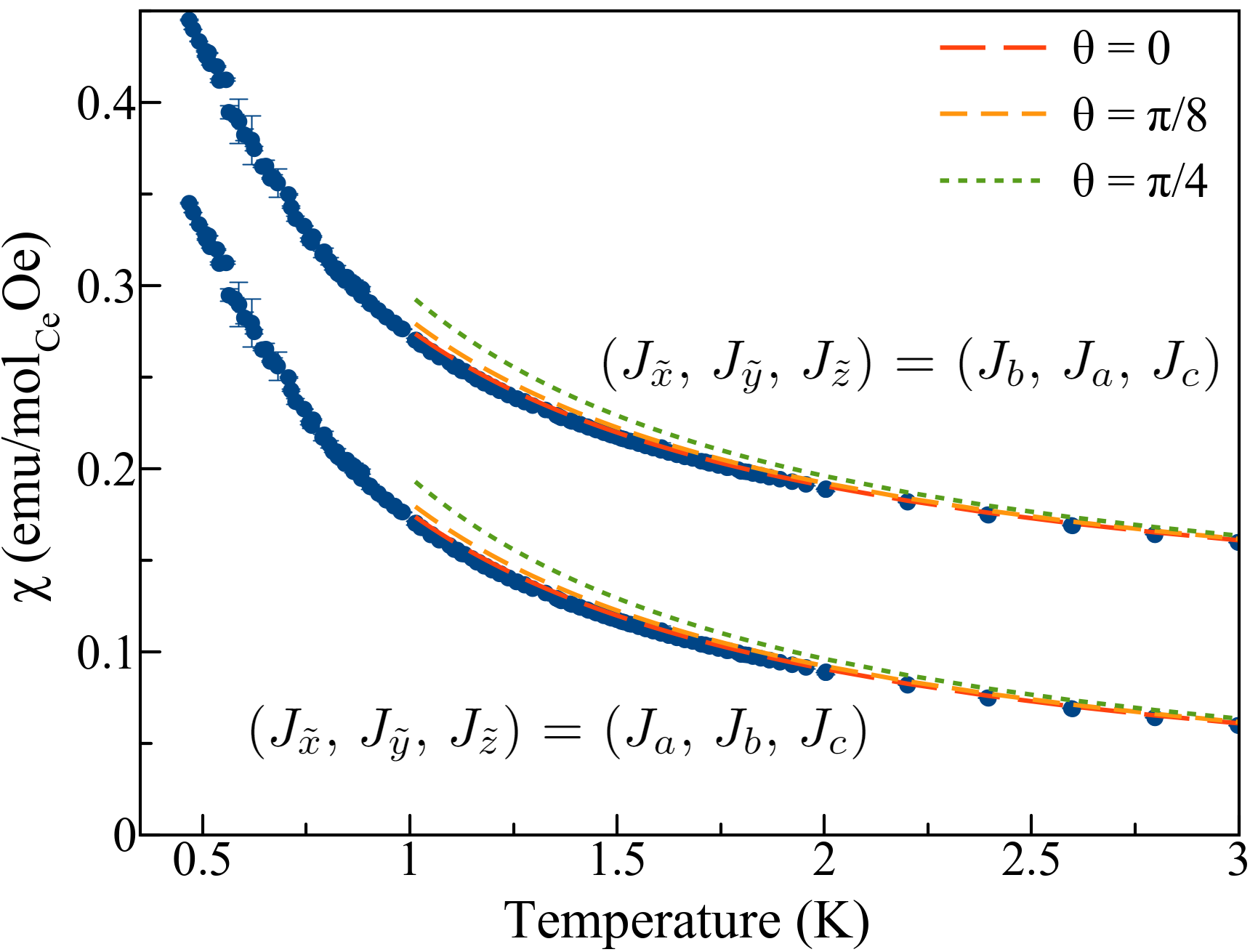}
\par
\caption{The measured powder magnetic susceptibility data is plotted alongside the 2$^{\mathrm{nd}}$ order NLC-calculated susceptibility for values of $\theta$ between 0 and $\pi/4$, and for $(J_{\tilde{x}}, J_{\tilde{y}}, J_{\tilde{z}})$ equal to the two permutations of the A parameters that are able to provide a reasonable fit to the data. Specifically, we show calculations for values of $\theta$ given by $\theta$ = 0 (red), $\theta$ = $\pi/8$ (yellow), and $\theta$ = $\pi/4$ (green). This shows that the NLC calculations for the magnetic susceptibility agree well with the data when $(J_{\tilde{x}}, J_{\tilde{y}}, J_{\tilde{z}})$ = $(0.064, 0.063, 0.011)$~meV, or $(J_{\tilde{x}}, J_{\tilde{y}}, J_{\tilde{z}})$ = $(0.063, 0.064, 0.011)$~meV, so long as the value of $\theta$ is near $\theta = 0$. The $(J_{\tilde{x}}, J_{\tilde{y}}, J_{\tilde{z}})$ = $(0.063, 0.064, 0.011)$~meV calculations are shifted upwards by 0.1 emu/mol$_{\mathrm{Ce}}$Oe for visibility.} 
\label{Figure7}
\end{figure}

\section{Consistency of Estimated Exchange Parameters with Neutron Scattering Results}

The combined analyses of the measured $C_\mathrm{mag}$ and $\chi$ give experimental estimates for the near neighbour exchange constants in Ce$_2$Zr$_2$O$_7$, yielding $\theta \sim 0$ and $(J_{\tilde{x}},  J_{\tilde{y}},  J_{\tilde{z}})$ = (0.064, 0.063, 0.011)~meV or $(J_{\tilde{x}},  J_{\tilde{y}},  J_{\tilde{z}})$ = (0.063, 0.064, 0.011)~meV. While neutron scattering measurements were {\it not} modelled in order to constrain the microscopic spin Hamiltonian for Ce$_2$Zr$_2$O$_7$, it is interesting and important to see to what extent the measured neutron scattering from Ce$_2$Zr$_2$O$_7$ is consistent with calculations using the near-neighbour spin Hamiltonian so derived.

\subsection{Elastic Neutron Scattering}

The U(1)$_\pi$ ground state, determined by these best-fitting near-neighbour exchange parameters, is consistent with the nature of the previously reported diffuse inelastic neutron scattering on single crystals of Ce$_2$Zr$_2$O$_7$~\cite{Gaudet2019, Gao2019}. Additionally, the earlier neutron scattering work is inconsistent with an ordered state, at least in the dipolar sector, as magnetic Bragg peaks would be expected. We have revisited our earlier elastic neutron scattering data to place an upper limit on possible All-In, All Out (AIAO) dipole order in the ground state of Ce$_2$Zr$_2$O$_7$, the form expected to reside within the XYZ DO pyrochlore phase diagram. We conclude that no such AIAO dipole order occurs in Ce$_2$Zr$_2$O$_7$, with an upper-limit on the Ce$^{3+}$ ordered moment of $\mu_\mathrm{ordered}$ $\leq 0.04$ $\mu_\mathrm{B}$ (see Appendix D).  

\subsection{Polarized Neutron Diffraction}

We can also compute the spin flip (SF) and non-spin flip (NSF) structure factors using this best-fitting A parameter set and compare with the polarized neutron diffraction measurements on an annealed single crystal sample of Ce$_2$Zr$_2$O$_7$ shown in Section III. The calculations are carried out at $T$ = 0.5~K (see Appendix E), as that is the lowest temperature for which the NLC-3 calculation converges, while the new polarized neutron diffraction measurements were performed at lower temperatures, $T$ = 0.045~K. Nonetheless, we assume that this calculation will capture most of the features at lower temperatures, as the ground state is disordered. 

The measured (NLC-calculated) SF scattering in the $[HHL]$ scattering plane is shown in Fig.~8(a) (8(b)) and the measured (NLC-calculated) NSF scattering in the $[HHL]$ scattering plane is shown in Fig.~8(c) (8(d)). The comparison between measurement and theory for the SF channel in Fig.~8(a) and 8(b) is good, although sharper features are present in the lower temperature, SF polarized diffraction, such as the broad pinch point scattering near (0,0,2). The measured NSF structure factor in the $[HHL]$ scattering plane (Fig.~8(c)) shows intensity that is maximal along Brillouin zone boundaries (shown as grey lines in Fig.~8(c)) and minimal at zone centres. As discussed in Section III, this zone boundary scattering is similar to that measured in the NSF channel of polarized neutron diffraction measurements on Ho$_2$Ti$_2$O$_7$, shown in Fig.~2(d) \cite{Fennell2009}, and associated with interactions beyond the nearest-neighbour. The calculated NSF structure factor is featureless for the near-neighbour-only XYZ spin Hamiltonian employed here, with a {\bf Q}-dependence originating from the Ce$^{3+}$ magnetic form factor only, as Fig.~8(d) illustrates. 

\begin{figure}[]
\linespread{1}
\par
\includegraphics[width=3.4in]{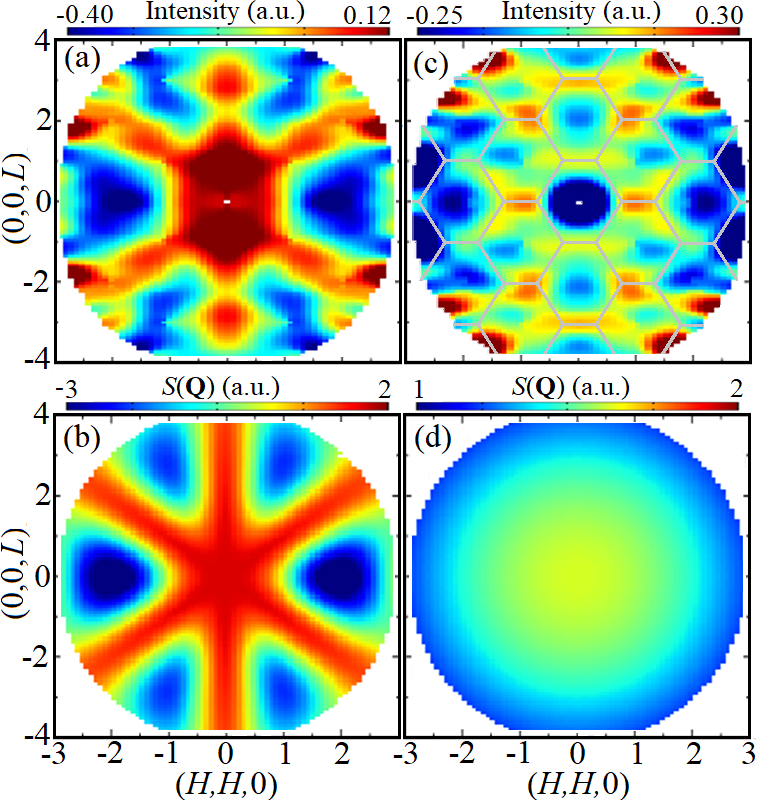}
\par
\caption{(a) The symmetrized $T$ = 45~mK - $T$ = 10~K temperature-difference neutron signal measured in the SF channel of our polarized neutron diffraction experiment. (b) The NLC-calculated equal-time structure factor for SF scattering in the $[HHL]$ plane at $T$ = 0.5~K with a $T$ = 10~K temperature subtraction. (c) The symmetrized $T$ = 45~mK - $T$ = 10~K temperature-difference neutron signal measured in the NSF channel of our polarized neutron diffraction experiment. The grey lines show the Brillouin zone boundaries. (d) The NLC-calculated equal-time structure factor for NSF scattering in the $[HHL]$ plane at $T$ = 0.5~K with a $T$ = 10~K temperature subtraction. Both (b) and (d) are calculated using the experimental estimates for the A near-neighbour exchange parameters yielded in this work (see main text).}
\label{Figure8}
\end{figure}

\subsection{Inelastic Neutron Scattering from Powder Samples}

Low energy, unpolarized inelastic neutron scattering measurements were performed on powder samples of Ce$_2$Zr$_2$O$_7$ as shown in Fig.~9(a), 9(b) and 9(c); this shows the temperature-difference neutron scattering spectra measured for a $T$ = 0.06~K, 0.5~K, and 3~K data-set with a $T$ = 9.6~K data-set used a background. This data was taken on the low energy disk chopper spectrometer (DCS) neutron instrument at NCNR with $E_{\mathrm{i}}$ = 3.27 meV incident neutrons giving an energy resolution of $\sim$0.09 meV at the elastic line. Inelastic data from this larger data set was previously discussed in Ref. \cite{Gaudet2019}.  We perform two sets of analysis with this data. First, in Fig.~10, we examine the temperature dependence of the measured and calculated integrated intensities for the $T$ = 9.6~K temperature subtraction, with integration in energy transfer over the range $E = [-0.2, 0.4]$~meV and integration in scattering vector over the range $|\mathbf{Q}| = [0.46, 0.93]~\angstrom^{-1}$. This integration range was chosen to enclose the dominant portion of the measured magnetic intensity, while avoiding nuclear Bragg peaks. The NLC calculations are carried out to 3$^{\mathrm{rd}}$ order (see Appendix F). For the A (B) exchange parameters, we use $\theta$ = 0 (0.561~radians), but it is important to note that there is no choice of $\theta$ for which the calculations using the B parameters agree with the temperature dependence of the experimental data over the range $|\mathbf{Q}| = [0.46, 0.93]~\angstrom^{-1}$.

We also compare these measurements with the corresponding spectra obtained via semi-classical Molecular Dynamics (MD) calculations based on Monte Carlo (MC) simulations (see Appendix H) using the near neighbour exchange parameters from the A regime, $(J_{\tilde{x}},  J_{\tilde{y}},  J_{\tilde{z}})$ = (0.064, 0.063, 0.011)~meV, for $\theta = 0$ (Fig.~9 (d, e, f)) and $\theta$ = $\pi/2$ (Fig.~9(g, h, i)). 

The temperature dependence of the measured signal is most consistent with that obtained from the semi-classical MD and MC simulations using $(J_{\tilde{x}},  J_{\tilde{y}},  J_{\tilde{z}})$ = (0.064, 0.063, 0.011)~meV when $\theta$ = 0. Furthermore, the energy dependence of the predicted signal is only consistent with the measured data for values of $\theta$ near $\theta = 0$. As $\theta$ increases from $\theta$ = 0 to $\theta$ = $\pi/2$, the spectral weight in the simulated signal shifts from $E \sim0.1$~meV to $E \sim0$~meV, as illustrated in Fig.~9(d, g). 

\begin{figure}[t]
\linespread{1}
\par
\includegraphics[width=3.4in]{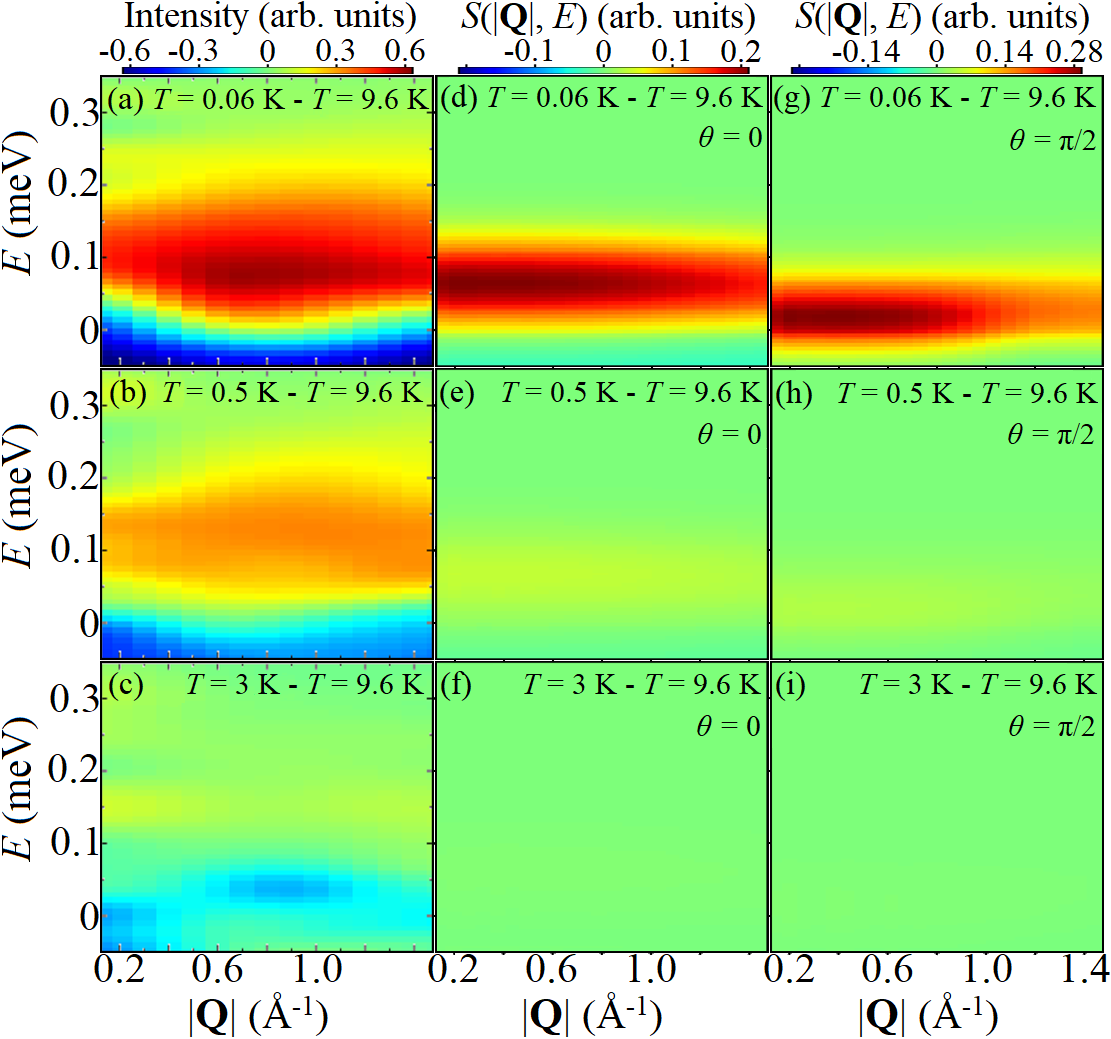}
\par
\caption{The measured inelastic neutron scattering from an annealed powder sample of Ce$_2$Zr$_2$O$_7$ is shown in (a), (b) and (c) for temperature-subtracted data relative to $T$ = 9.6~K.  The corresponding powder-averaged neutron scattering structure factor ($S(\lvert\mathbf{Q}\rvert, E, T)$) calculated from semi-classical Molecular Dynamics calculations based on Monte Carlo simulations using near-neighbour exchange parameters from the A regime, $(J_{\tilde{x}},  J_{\tilde{y}},  J_{\tilde{z}})$ = (0.064, 0.063, 0.011)~meV, are shown in panels (d) through (i). The temperatures of the measured and calculated data sets (T=0.06 K, 0.5 K and 3 K) as well as the $\theta$ values used in the calculations are as indicated in the individual panels.}
\label{Figure9}
\end{figure}

\begin{figure}[t]
\linespread{1}
\par
\includegraphics[width=3.4in]{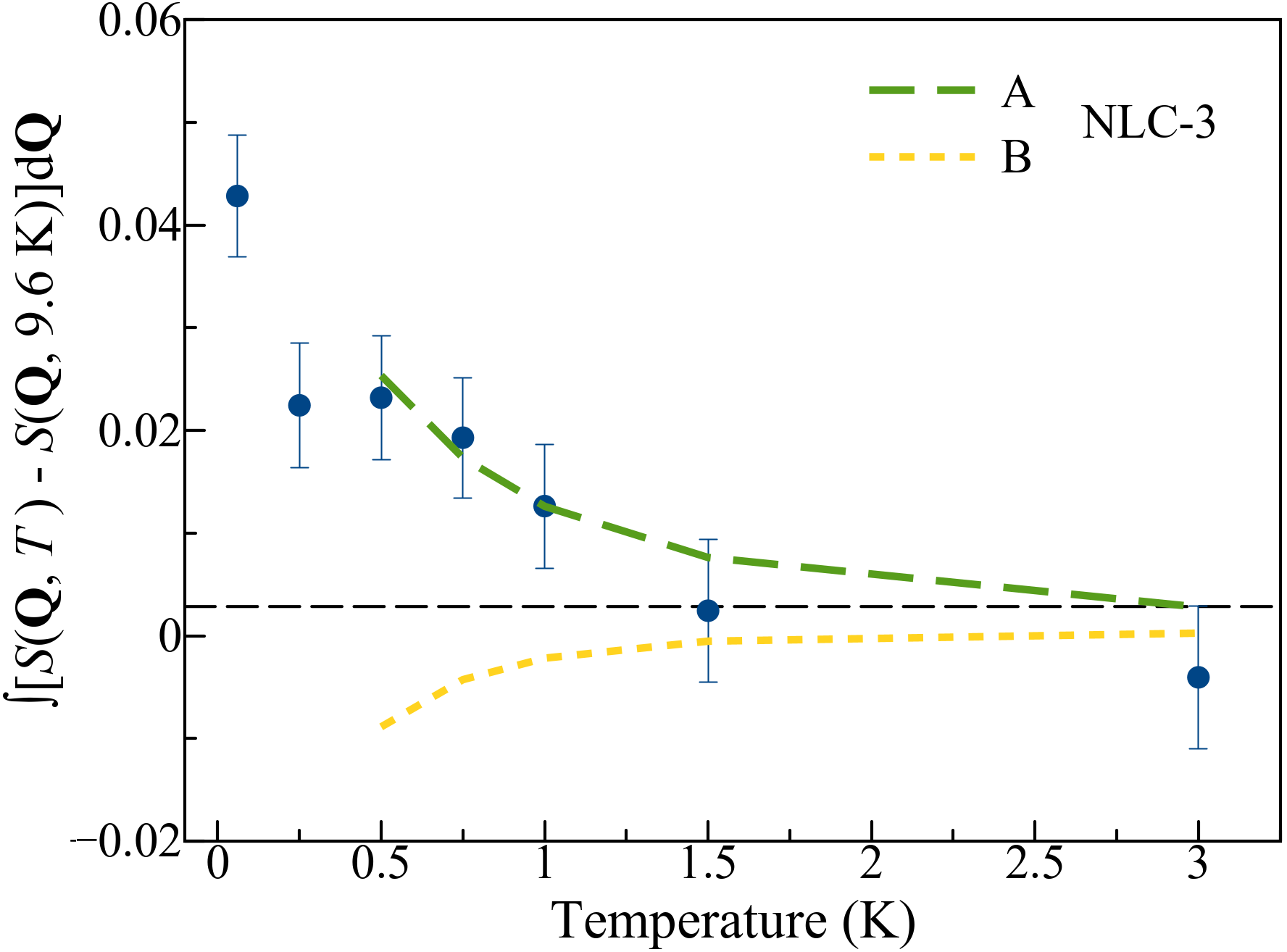}
\par
\caption{The results of the NLC $S(\mathbf{Q},T)$-calculation to 3$^{\mathrm{rd}}$ order using the A and B exchange parameters, overlaid on top of the measured neutron scattering intensity from our Ce$_2$Zr$_2$O$_7$ sample. Here we compare the temperature dependence of the measured and calculated integrated intensities for the $T$ = 9.6~K temperature subtraction, with integration over the energy-transfer range $E = [-0.2, 0.4]$~meV and integration in wavevector over the range $|\mathbf{Q}| = [0.46, 0.93]~\angstrom^{-1}$. The temperature dependence of the NLC-calculated integrated-$S(\mathbf{Q},T)$ agrees well with that of the measured data when using parameter set A, but clearly does not for set B.}
\label{Figure10} 
\end{figure}

\section{Discussion}

\subsection{Low Temperature Heat Capacity and Entropy}

The new $C_p$ measurements also provide better definition of the low temperature $C_\mathrm{mag}$, below $\sim$0.1~K, where $C_\mathrm{mag}$ falls off sharply towards zero. The lowest-temperature data points can be used to model how $C_\mathrm{mag}$ approaches zero at $T$ = 0~K. This is interesting to do because an extrapolation of $C_\mathrm{mag}$ below experimentally accessible temperatures to $T$ = 0~K, allows us to evaluate the entropy $S_\mathrm{mag}(T) = \int_{0}^T \frac{C_\mathrm{mag}}{T} dT$.  

The two simple forms for the low temperature $C_\mathrm{mag}$, an exponential form and a cubic form, are shown in the inset of Fig.~3.  Both forms are too simple to be related to the spin Hamiltonian or U(1)$_\pi$ ground state in any sophisticated manner; however one can smoothly extrapolate the low temperature $C_\mathrm{mag}$ data to zero using an exponentially activated form. A simple power law, such as the cubic form in the inset of Fig.~3, does not smoothly meet up with the low temperature data at the lowest measured temperature, $T$ = 0.058~K; doing so would require a non-physical sub-linear $C_\mathrm{mag}$ at the lowest temperatures. A cubic extrapolation was used in the previous work on the $C_\mathrm{mag}$ of Ce$_2$Zr$_2$O$_7$ (Ref.~\cite{Gao2019}), however our new results, consistent with the previous measurements, show that such a low temperature extrapolation is inappropriate. 

The cubic form would be appropriate for emergent gapless photon excitations associated with U(1) QSLs~\cite{Li2017, Kato2015, Huang2018b}. However, depending on the speed of light for these emergent photons, their $T^3$ contribution may only enter at very low temperatures~\cite{Benton2012}.  Furthermore the bending of the photon dispersion towards the zone boundary, combined with contributions from gapped spinons and visons, can easily mimic the exponentially-activated form at intermediate temperatures. Interactions between visons and photons can also cause the photons to develop an effective temperature dependent gap \cite{Kwasigroch2020}. To address these subtleties, we use a low $T$ form for $C_\mathrm{mag}$ which is based on an interpolation scheme connecting the $T$ $>$ $\sim$0.5~K $C_\mathrm{mag}$ regime described by the NLC calculations, and hence consistent with the proposed spin Hamiltonian, to a low temperature form consistent with a $T^3$ $C_\mathrm{mag}$ from U(1) emergent photons at sufficiently low temperatures. This involves an interpolation scheme for $C_\mathrm{mag}$ and S$_\mathrm{mag}$ following the method of Pade approximants in Ref.~\cite{Bernu2001} (see Appendix I).  The resulting theoretical curve, now covering all temperatures, is shown as the solid line in Fig.~11(a,~b). Clearly the low temperature portion of this curve smoothly connects to the low temperature $C_\mathrm{mag}$ data. The point of this exercise is to provide a physically-motivated form of $C_\mathrm{mag}$ which extrapolates smoothly between the lowest temperature $C_\mathrm{mag}$ data point and zero at $T$ = 0~K.

With a good minimal description of $C_\mathrm{mag}$ for Ce$_2$Zr$_2$O$_7$ at the lowest temperatures in place, we can look to account for the entropy associated with the DO doublet, which must be $R\ln(2)$, as this ground state doublet is well separated, by $\sim$55~meV, from the first excited CEF state~\cite{Gaudet2019, Gao2019}. Fig.~11(b) shows the integration of the $C_\mathrm{mag}/T$ data to give the entropy $S_\mathrm{mag}$ to $\sim$10~K. The experimental entropy of $R\ln(2)$ is recovered, to within 5$\%$ which may be associated with 4\textit{f}$^0$ Ce$^{4+}$ impurities. Interestingly, Fig.~11(b) also shows that accounting for the entropy from the only feature in the temperature dependence of $C_\mathrm{mag}$, the beginning of the $C_\mathrm{mag}$ plateau at $T$ = 0.08~K, to 10~K gives $\sim$$R[\ln(2) - \frac{1}{2}\ln(\frac{3}{2})]$, the Pauling entropy associated with both spin ice and proton disorder in solid ice. Note that this latter argument is independent of the low temperature extrapolation of $C_\mathrm{mag}$.

\begin{figure}[t]
\linespread{1}
\par
\includegraphics[width=3.4in]{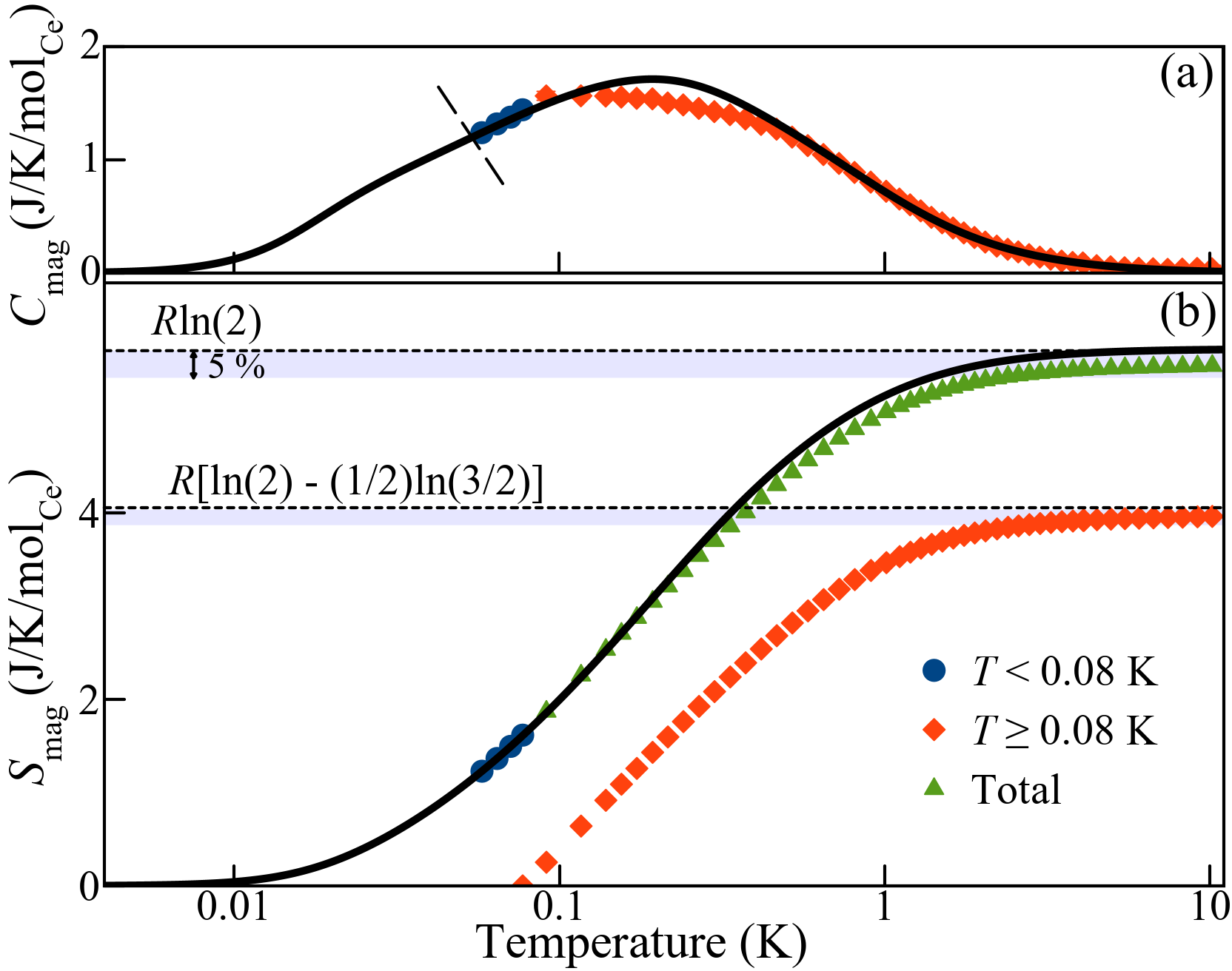}
\par
\caption{ (a) The measured $C_\mathrm{mag}$ and best-fit $C_\mathrm{mag}$-interpolation for the Ce$_2$Zr$_2$O$_7$ sample of the present work. The data is divided into high and low $T$ regimes around $T$ = 0.08~K, which separates the plateau regime from the rapidly decreasing $C_\mathrm{mag}$ regime. (b) The magnetic entropy recovered from $S_\mathrm{mag} = \int_{T_0}^T \frac{C_\mathrm{mag}}{T} dT $ over the full temperature range (T$_0$ = 0~K) and above the onset of the plateau (T$_0$ = 0.08~K) are shown. This is derived from the integration of $C_\mathrm{mag}$ shown in (a), and employs the $C_\mathrm{mag}$-interpolation below the lowest measured temperature, accounting for gapless photons as well as gapped spinons and visons. $R\ln(2)$ in entropy is recovered over the full temperature range, to within 5$\%$, which is the approximate deficiency expected for Ce$^{4+}$ in this sample. The Pauling spin ice entropy $R[\ln(2) - \frac{1}{2}\ln(\frac{3}{2}$)] is recovered from the onset of the plateau, $T$ = 0.08~K, to $T$ = 10~K to within approximately the same tolerance.}
\label{Figure11}
\end{figure}

\subsection{Implications of Small $\theta$}
 
In the case where $J_{\tilde{x}}$ is the largest exchange parameter in the XYZ Hamiltonian, the resulting U(1)$_{\pi}$ QSL is dipolar from a symmetry perspective. Its emergent electric field transforms like a magnetic dipole. However, the small value of $\theta$ suppresses coupling between the emergent field and external magnetic fields. Therefore, for this case, we expect weak coupling between neutrons and emergent photons at low $|\mathbf{Q}|$. In the case of $J_{\tilde{y}}$ $>$ $J_{\tilde{x}}$, there would be no low-$\lvert\textbf{Q}\rvert$ coupling between photons and neutrons regardless of the value of $\theta$. It is therefore unlikely that the inelastic neutron scattering signal observed at low energy in Refs.~\cite{Gaudet2019, Gao2019} (and in this work) originates from an integration over emergent photons, despite the similarity to predictions in~\cite{Benton2012}. The dominant neutron scattering signal should then come from gapped spinons. 

A further implication of the small value for $\theta$ is that spin waves in finite magnetic field will be difficult to observe.  This may be important to note as modelling spin wave dispersion and intensity in a field-polarized state has been effectively applied to understanding the microscopic ground state in several pyrochlore magnets based on Kramer's doublet CEF ground states~\cite{Ross2011, Ross2014, Scheie2020, Savary2012, Thompson2017, Robert2015}. It may also underlie the lack of observation of well defined spin waves in studies of Ce$_2$Zr$_2$O$_7$ published to date.  A finite value of $\theta$ implies that the local magnetic moment operator possesses components transverse to the expectation value of the pseudospins in the high field state.  It is the finite transverse matrix elements which allow the observation of single spin waves by inelastic neutron scattering. In contrast, when $\theta$ = 0, the magnetic moment operator is parallel to the pseudospin directions in the high field state, and the matrix element connecting the ground state to single spin wave excitations is zero.

\section{Summary and Conclusions}

To conclude, we report new spin polarized neutron diffraction and $C_\mathrm{mag}$ measurements on single crystal Ce$_2$Zr$_2$O$_7$ in zero magnetic field. Our modelling of $C_\mathrm{mag}$, $\chi$, and $S(\mathbf{Q},T)$ with NLC calculations provides strong constraints on the exchange terms in the microscopic near neighbour XYZ Hamiltonian.  We arrive at best fit Hamiltonian parameters $\theta \sim 0$ and $(J_{\tilde{x}},  J_{\tilde{y}},  J_{\tilde{z}})$ = (0.064, 0.063, 0.011)~meV or $(J_{\tilde{x}},  J_{\tilde{y}},  J_{\tilde{z}})$ = (0.063, 0.064, 0.011)~meV, which indicates that a U(1)$_\pi$ QSL ground state is selected near the boundary between dipolar and octupolar character. 

The best-fitting exchange parameters from this work largely describe the SF neutron diffraction signal measured from single crystal Ce$_2$Zr$_2$O$_7$, while zone boundary scattering in the NSF channel indicates the significance of interactions beyond near-neighbour, including long-ranged dipolar interactions. The 7$^{\mathrm{th}}$ order NLC calculations for $C_\mathrm{mag}$ evaluated at the best fit Hamiltonian parameters do not describe the measured $C_\mathrm{mag}$ at the lowest temperatures, again consistent with weak interactions in Ce$_2$Zr$_2$O$_7$'s Hamiltonian beyond near-neighbour and beyond the scope of the present calculations.  

The new $C_\mathrm{mag}$ data extends to temperatures as low as $T$ = 0.058~K and can be smoothly extrapolated to zero temperature using a form consistent with both the XYZ spin Hamiltonian estimated from fitting the NLC calculations to the data, and with a $T^3$ form for $C_\mathrm{mag}$ at sufficiently low temperatures, appropriate to emergent gapless photons. With such a low $T$ form for $C_\mathrm{mag}$ in place we show the $R\ln(2)$ entropy associated with Ce$^{3+}$'s DO doublet ground state is recovered to 10~K. Phenomenologically, we observe that the Pauling entropy for spin ice is recovered above the onset of the $T$ $\sim$0.08~K plateau in $C_\mathrm{mag}$. 

\begin{acknowledgments}
We greatly appreciate the technical support from Alan Ye and Yegor Vekhov at the NIST Center for Neutron Research. This work was supported by the Natural Sciences and Engineering Research Council of Canada (NSERC). We also acknowledge the support of the National Institute of Standards and Technology, U.S. Department of Commerce. Certain commercial equipment, instruments, or materials (or suppliers, or software, etc) are identified in this paper to foster understanding. Such identification does not imply recommendation or endorsement by the National Institute of Standards and Technology, nor does it imply that the materials or equipment identified are necessarily the best available for the purpose. D.R.Y. and K.A.R. acknowledge the use of the Analytical Resources Core at Colorado State University. B.P. and R.S. acknowledge support by the Deutsche Forschungsgemeinschaft under grants SFB 1143 (project-id 247310070) and the cluster of excellence ct.qmat (EXC 2147, project-id 390858490). We thank Paul McClarty for a critical reading of the manuscript.

\end{acknowledgments}


\section*{Appendix A: Synthesis and Characterization}

The powder and single crystal samples of Ce$_2$Zr$_2$O$_{7}$ used in this work were prepared and characterized as described in Ref.~\cite{Gaudet2019}. La$_2$Zr$_2$O$_{7}$ was synthesized in order to estimate the phonon contribution to the $C_p$ of Ce$_2$Zr$_2$O$_{7}$. The powder samples of La$_2$Zr$_2$O$_{7}$ measured in this work were first prepared by mixing stoichiometric amounts of La$_2$O$_3$ (Alfa Aesar 99.99\%) and ZrO$_2$ (Alfa Aesar 99.7\%). The La$_2$O$_3$ (ZrO$_2$) powder was precalcined (dried) at 800$^{\circ}$C (200$^{\circ}$C) prior to mixing. The stoichiometric mixture was pelletized and sintered in air at 1350$^{\circ}$C for 36 hours, three times, with regrinding and repelletization between sinterings. Fig.~12 shows an x-ray Rietveld refinement against the $Fd\bar{3}m$ space group for a typical powder sample of La$_2$Zr$_2$O$_{7}$ synthesized for this work. 

\section*{Appendix B: Heat Capacity and Numerical Linked Cluster Calculations with 5\% Oxidation}

We provide further details on the results of our 4$^{\mathrm{th}}$-order NLC calculations for $C_\mathrm{mag}$ with a 5\% oxidation level included in the calculations. The calculated $C_\mathrm{mag}$ with 5\% oxidation using the globally (locally) best-fitting exchange parameters that are also able to describe the measured susceptibility, A' (B'), is shown in Fig.~13(a) (13(b)). To improve convergence of the NLC calculations, we have used the Euler transformation to the 3$^{\mathrm{rd}}$ (Euler 3) and 4$^{\mathrm{th}}$ (Euler 4) orders (see Appendix J). While the parameter sets A’ and B’ are both locally optimal, the A’ description of the $C_\mathrm{mag}$ data is clearly superior. 

The inset to Fig.~13(b) shows the two locally optimal regions of parameter space for the 4$^{\mathrm{th}}$-order NLC calculations of $C_\mathrm{mag}$, for both a 0\% oxidation level and a 5\% oxidation level, defined by $\log \langle \frac{\delta^2}{\epsilon^2} \rangle_{C\mathrm{mag}} < 2.7$ for the purposes of the visualization. From the similarity of these regions and their local minima (A and A', B and B'), we conclude that the results of the NLC calculations for $C_\mathrm{mag}$ are robust to the sample oxidation up to oxidation levels of at least 5\%. In Table~1 we summarize the results of our NLC fittings to $C_\mathrm{mag}$ and list the best-fitting exchange parameters corresponding to each fitting.

\begin{figure}
\linespread{1}
\par
    \begin{center}
        \includegraphics[width=3.4in]{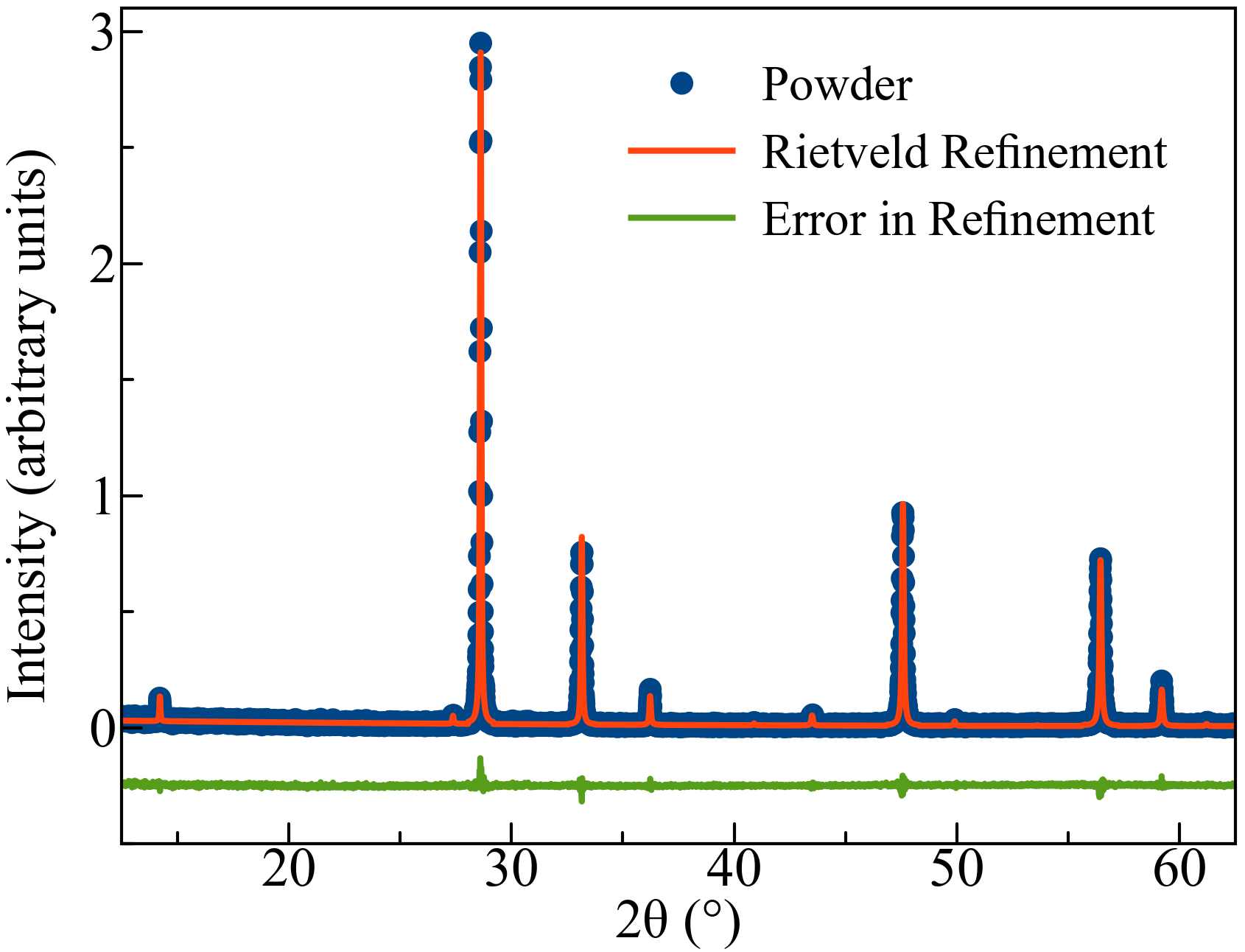}
        \par
        \caption{Powder x-ray refinement of the La$_2$Zr$_2$O$_7$ sample synthesized for this work. The difference between the measured and calculated diffraction patterns is shown in green and indicates phase purity; this line has been shifted downwards by 0.25 units for visibility.} 
        \end{center}
\label{Figure12}
\end{figure}

\begin{table}[h]
\label{ExchangeConstants}
\begin{tabular}{|c|c|c|c|c|c|}
\hline
Set & Oxidation & $J_a$(meV) & $J_b$(meV) & $J_c$(meV)\\\hline

\begin{tabular}[c]{@{}c@{}} A \end{tabular}          
& 0\% & 0.064 & 0.063 & 0.011 \\ \hline

\begin{tabular}[c]{@{}c@{}} A' \end{tabular}          
& 5\% & 0.067 & 0.067 & 0.012 \\ \hline

\begin{tabular}[c]{@{}c@{}} B \end{tabular}          
& 0\% & 0.089 & -0.007 & -0.027 \\ \hline

\begin{tabular}[c]{@{}c@{}} B' \end{tabular}          
& 5\% & 0.089 & 0.006 & -0.037 \\ \hline

\end{tabular}
\caption{A summary of the different sets of near-neighbour exchange constants discussed throughout this work. Each set of exchange constants was determined according to the minimization of the goodness-of-fit parameter, $\langle \frac{\delta^2}{\epsilon^2} \rangle_{C\mathrm{mag}}$, corresponding to 4$^{\mathrm{th}}$-order NLC calculations for $C_\mathrm{mag}$ with a low temperature threshold of $0.7J_a/k_\mathrm{B}$ in the evaluation of $\langle \frac{\delta^2}{\epsilon^2} \rangle_{C\mathrm{mag}}$. We also list the level of sample oxidation considered in each calculation.}
\end{table}

\begin{figure*}
\linespread{1}
\par
    \begin{center}
        \includegraphics[width=7in]{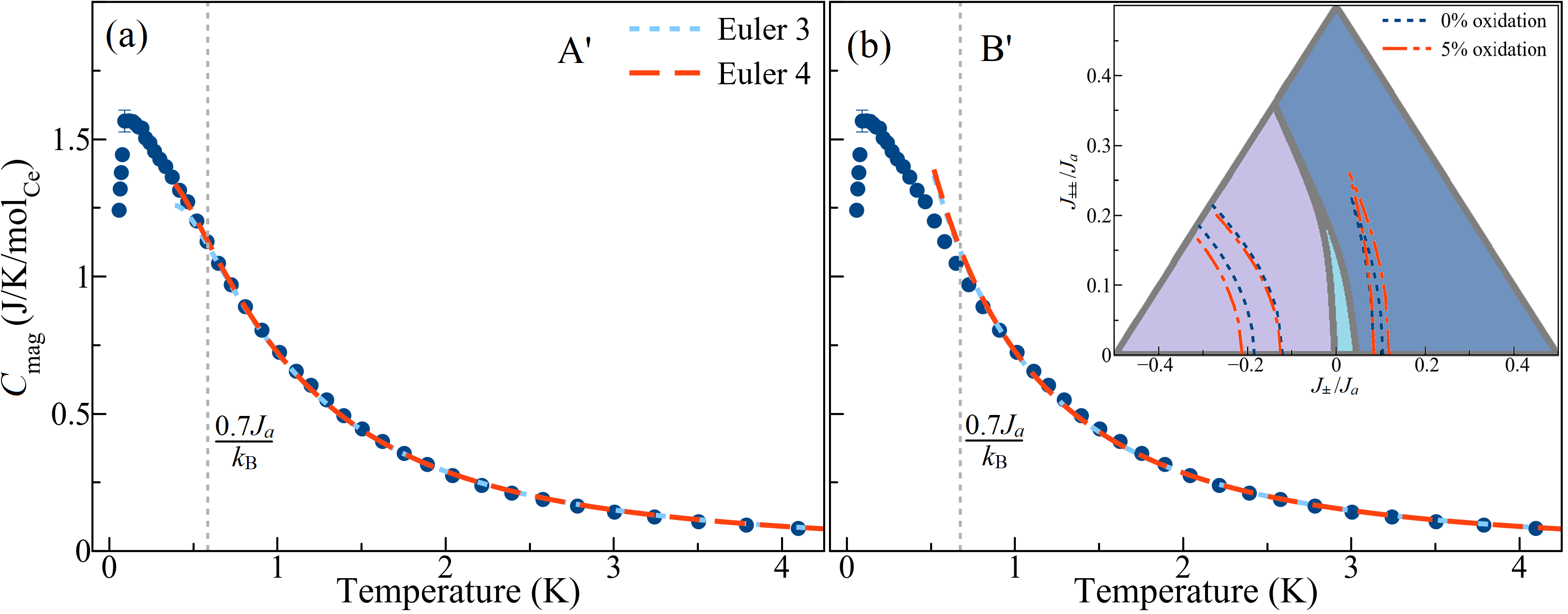}
        \par
        \caption{(a) The results of the 4$^{\mathrm{th}}$ order NLC $C_\mathrm{mag}$-calculation for 5\% sample-oxidation, using the near-neighbour exchange parameters $J_a$ = 0.067~meV, $J_b$ = 0.067~meV, and $J_c$ = 0.012~meV (set A'), overlaid on top of the measured $C_\mathrm{mag}$ for our Ce$_2$Zr$_2$O$_7$ sample. (b) The results of the 4$^{\mathrm{th}}$ order NLC $C_\mathrm{mag}$-calculation for 5\% sample-oxidation, using the near-neighbour exchange parameters $J_a$ = 0.089~meV, $J_b$ = 0.006~meV, and $J_c$ = -0.037~meV (set B'), overlaid on top of the measured $C_\mathrm{mag}$ for our Ce$_2$Zr$_2$O$_7$ sample. We have used Euler transformations to the 3$^{\mathrm{rd}}$ (Euler 3) and 4$^{\mathrm{th}}$ (Euler 4) orders to improve convergence of the NLC $C_\mathrm{mag}$-calculations (see Appendix J). The inset to (b) shows a comparison of the locally optimal fitting-regions obtained from NLC calculations with an oxidation level of 0\% (blue) and 5\% (red). For visualization purposes, the optimal fitting regions in this plot are defined by $\log \langle \frac{\delta^2}{\epsilon^2} \rangle_{C\mathrm{mag}} < 2.7$, where $\langle \frac{\delta^2}{\epsilon^2} \rangle_{C\mathrm{mag}}$ is the goodness-of-fit parameter for the NLC calculations as described in the main text. We overplot this on the predicted ground state phase diagram for the XYZ model Hamiltonian~\cite{Benton2020}, but omit the labels for aesthetic purposes (see Fig.~4(b) for labels). Conclusions from fitting the NLC calculations to the data are robust to a 5\% oxidation level.} 
        \end{center}
\label{Figure13}
\end{figure*}

\section*{Appendix C: NLC Fitting to $\chi$}

In this section we discuss the results of the NLC-fitting to the magnetic susceptibility measured from a powder sample of Ce$_2$Zr$_2$O$_7$. The magnetic susceptibility is dependent on the values of $J_{\tilde{x}}$, $J_{\tilde{y}}$, $J_{\tilde{z}}$, and $\theta$. The exchange parameters $(J_{\tilde{x}},  J_{\tilde{y}},  J_{\tilde{z}})$, are given by some permutation of $(J_a,  J_b,  J_c)$. We allow $\theta$ to vary in the range from 0 to $\pi$/4. This is enough to cover all distinguishable scenarios, since changing the sign of $\theta$ does not affect any quantity considered here, and shifting $\theta$ to $\theta + \pi$/2 is the same as reversing the sign of $\theta$ and swapping the values of $J_{\tilde{x}}$ and $J_{\tilde{z}}$, which is already covered by considering all six permutations of exchange parameters.

NLC calculations up to 2$^{\mathrm{nd}}$ order were performed to compute the powder-averaged magnetic susceptibility and to compare the calculations to the corresponding measurement on Ce$_2$Zr$_2$O$_7$. A constant term was added to the NLC calculations to account for the effect of mixing in higher crystal field levels due to an applied magnetic field. This term is calculated from the low-temperature limit of single ion susceptibility using the crystal-field scheme of Ce$^{3+}$ in Ce$_2$Zr$_2$O$_7$ reported in Ref.~\cite{Gaudet2019}. The level of sample oxidation for the measured powder sample had an upper limit of $\sim$14\%. This upper limit was estimated from fits to the single ion susceptibility at high temperature using the crystal-field scheme of Ce$^{3+}$ in Ce$_2$Zr$_2$O$_7$ reported in Ref.~\cite{Gaudet2019}. Accordingly, a 14\% oxidation level is included in our NLC calculations of the magnetic susceptibility. 

NLC calculations of the magnetic susceptibility were performed for parameter sets throughout the A and B regions identified by the $C_\mathrm{mag}$ fittings. The calculations were compared with experimental data between 1~K and 10~K. Fig.~5 shows the regions of the phase diagram for which it is possible to obtain simultaneous agreement with $C_\mathrm{mag}$ and $\chi$, for $(J_{\tilde{x}},  J_{\tilde{y}},  J_{\tilde{z}})$ equal to the different permutations of $(J_a,  J_b,  J_c)$. We define these regions by the simultaneous satisfaction of $\log \langle \frac{\delta^2}{\epsilon^2} \rangle_{C\mathrm{mag}} < 2.7$ and $\log \langle \frac{\delta^2}{\epsilon^2} \rangle_{\chi} < 12.1$. The goodness-of-fit measure $\langle \frac{\delta^2}{\epsilon^2} \rangle_{C\mathrm{mag}}$ is defined in the main text, and $\langle \frac{\delta^2}{\epsilon^2} \rangle_{\chi} \propto \sum\frac{(\chi^\mathrm{NLC}(T_\mathrm{exp})-\chi^\mathrm{exp}(T_\mathrm{exp}))^2}{\epsilon(T_\mathrm{exp})^2}$, where the sum is over measured temperatures $T_\mathrm{exp}$ between 1 K and 10 K and $\epsilon(T_\mathrm{exp})$ is the experimental uncertainty on the magnetic susceptibility at temperature $T_{exp}$. We allow $\theta$ to vary in the range from 0 to $\pi$/4 in finding the best agreement with the susceptibility data for each permutation. The relatively small experimental uncertainties on the magnetic susceptibility contribute to the larger upper limit for $\langle \frac{\delta^2}{\epsilon^2} \rangle_{\chi}$ in comparison to the upper limit used for $\langle \frac{\delta^2}{\epsilon^2} \rangle_{C\mathrm{mag}}$. 

\section*{Appendix D: Elastic Neutron Scattering}

\begin{figure*}
\linespread{1}
\par
    \begin{center}
        \includegraphics[width=7in]{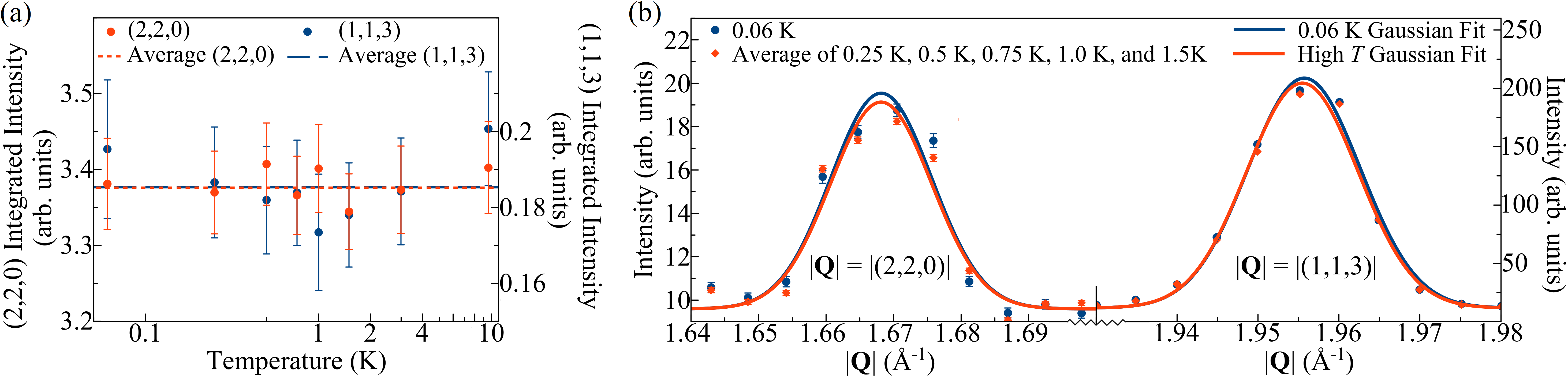}
        \par
        \caption{(a) The temperature dependence of the integrated intensity for the Bragg peaks at the $\textbf{Q} = (2,2,0)$ (red) and $\textbf{Q} = (1,1,3)$ (blue) positions. No significant temperature dependence is discernible. (b) Elastic Q-cuts through the $\textbf{Q} = (2,2,0)$ (left) and $\textbf{Q} = (1,1,3)$ (right) positions at $T$ = 0.06~K (blue) and averaged over the higher temperature data points $T$ = 0.25~K, 0.5~K, 0.75~K, 1~K, 1.5~K (red). The Gaussian fitting to each of these data sets, used to determine a corresponding integrated intensity, is also shown for each Bragg peak in (b). From these integrated intensities, we conclude that no AIAO dipole order occurs in Ce$_2$Zr$_2$O$_7$, with an upper-limit on the Ce$^{3+}$ ordered moment of $\mu_\mathrm{ordered}$ $\leq 0.04$ $\mu_\mathrm{B}$.} 
        \end{center}
\label{Figure14}
\end{figure*}

In this section we discuss the analysis of our elastic neutron scattering data, measured on an annealed powder sample of Ce$_2$Zr$_2$O$_7$ and used to place an upper limit of $\mu_\mathrm{ordered}$ $\leq 0.04$ $\mu_\mathrm{B}$ on the ordered moment corresponding to any All-in, All-out (AIAO) dipole order in Ce$_2$Zr$_2$O$_7$'s magnetic ground state. The strongest magnetic Bragg peaks associated with AIAO order are expected to reside at the $\textbf{Q} = (2,2,0)$ and $\textbf{Q} = (1,1,3)$ positions of reciprocal space. Accordingly, we can examine the temperature dependence of the scattered intensity at these locations in order to look for increases of intensity with decreasing temperature, which would signal the onset of a magnetic Bragg peak and associated magnetic order. As shown in Fig.~14(a), no such increase in intensity is detected upon lowering temperature. 

In Fig.~14(b), we show the measured intensity around the $\textbf{Q} = (2,2,0)$ (left) and $\textbf{Q} = (1,1,3)$ (right) positions at $T$ = 0.06~K (blue) and as averaged over the temperatures $T$ = 0.25~K, $T$ = 0.5~K, $T$ = 0.75~K, $T$ = 1~K, and $T$ = 1.5~K (red). Gaussian fits to the peak at each of the locations are also shown for each temperature (or temperature-average) and the area underneath of these Gaussian peaks was used in order to determine the corresponding integrated intensity for each peak. From these values for the integrated intensity, we place an upper limit of $\mu_\mathrm{ordered}$ $\leq 0.04$ $\mu_\mathrm{B}$ for the Ce$^{3+}$ ordered moment corresponding to any AIAO magnetic dipole ordering in Ce$_2$Zr$_2$O$_7$. 

For each selected Bragg peak position $\textbf{Q}$, an upper limit is calculated in accordance with the equation, 
\begin{equation}\tag{D1}
\mu_\mathrm{ordered} = \Bigg(\frac{I^\mathrm{exp}_\mathrm{mag}}{{{I}^\mathrm{exp}_\mathrm{nuc}}}\Bigg)^{\frac{1}{2}} \frac{|F(\textbf{Q})|}{|\mathbf{F^\mathrm{mag}_{\perp}(\textbf{Q})}/\mu|} ,
\end{equation}
where ${I}^\mathrm{exp}_\mathrm{mag}$ and ${I}^\mathrm{exp}_\mathrm{nuc}$ are the measured magnetic and nuclear contributions to the integrated Bragg intensity, respectively. $F(\textbf{Q})$ is the nuclear structure factor and $\mathbf{F^\mathrm{mag}_{\perp}}(\textbf{Q})/\mu$ is the component of the magnetic structure factor that is perpendicular to $\textbf{Q}$, after dividing out the magnitude of the ordered moment ($\mu$) from the calculation~\cite{Skold1986}.  

\section*{Appendix E: Polarized Neutron Scattering Measurements and Calculations}

We have used 3$^{\mathrm{rd}}$ order NLC calculations to compute the energy-integrated scattering signals corresponding to a polarized neutron scattering experiment with sample alignment in the $[HHL]$ scattering plane. The exchange parameters are set to $\theta = 0$ and $(J_{\tilde{x}},  J_{\tilde{y}},  J_{\tilde{z}})$ = (0.064, 0.063, 0.011)~meV for the calculation and we perform the NLC-3 calculation with $T$ = 0.5~K as that is the lowest temperature for which the NLC expansion converges. Specifically, we compute,

\begin{equation}\tag{E1}
\begin{split}
    S^\mathrm{SF}({\bf Q}) = \frac{1}{N} |f(|{\bf Q}|)|^2 \sum_{ij} \Big[ e^{i {\bf Q} \cdot ({\bf r}_i - {\bf r}_j)}  (\hat{{\bf u}}({\bf Q}) \cdot \hat{{\bf z}}_i) (\hat{{\bf u}}({\bf Q}) \cdot \hat{{\bf z}}_j)&\\ 
    \left(\sin^2(\theta) \langle S^{\tilde{x}}_i S^{\tilde{x}}_j \rangle + \cos^2(\theta) \langle S^{\tilde{z}}_i S^{\tilde{z}}_j\rangle\right) \Big] &
\end{split}
\end{equation}

and,

\begin{equation}\tag{E2}
\begin{split}
    S^\mathrm{NSF}({\bf Q}) = \frac{1}{N} |f(|{\bf Q}|)|^2 \sum_{ij} \Big[ e^{i {\bf Q} \cdot ({\bf r}_i - {\bf r}_j)}  (\hat{{\bf n}} \cdot \hat{{\bf z}}_i) (\hat{{\bf n}} \cdot \hat{{\bf z}}_j)&\\ 
    \left(\sin^2(\theta) \langle S^{\tilde{x}}_i S^{\tilde{x}}_j \rangle + \cos^2(\theta) \langle S^{\tilde{z}}_i S^{\tilde{z}}_j\rangle\right) \Big] &
\end{split}
\end{equation}

where $S^\mathrm{SF}(\mathbf{Q}))$ ($S^\mathrm{NSF}(\mathbf{Q})$) denotes the energy-integrated structure factor for SF (NSF) scattering. $N$ is the number of spins in the lattice, $f(|{\bf Q}|)$ is the magnetic form factor for Ce$^{3+}$ (calculated using the analytical approximation in Ref. \cite{Lisher1971}), $\hat{{\bf n}}$ is the neutron polarisation direction, $\hat{{\bf z}}_i$ is the local anisotropy axis for the site $i$ and,

\begin{equation}\tag{E3}
\hat{{\bf u}}({\bf Q}) = \frac{   \hat{{\bf n}}  \times {\bf Q} }{|  \hat{{\bf n}}  \times {\bf Q}  |}.
\end{equation}

We compute $S^\mathrm{SF}(\mathbf{Q}))$ and $S^\mathrm{NSF}(\mathbf{Q})$ at $T$ = 0.5~K and in each case we subtract the corresponding calculation at $T$ = 10~K for better comparison with the temperature-subtracted experimental data. In Fig.~8(a, b) (Fig.~8(c, d)) of the main text, we compare the NLC-calculated $S^\mathrm{SF}(\mathbf{Q})$ ($S^\mathrm{NSF}(\mathbf{Q})$) to polarized neutron scattering measurements performed on an annealed $\sim$1.5 g single crystal sample of Ce$_2$Zr$_2$O$_7$ using the D7 diffractometer at the Institut Laue-Langevin with an incident energy of $E_i$ = 3.47~meV and a dilution refrigerator sample environment. The sample was aligned in a copper sample holder in the $[HHL]$ scattering plane with the uniaxial polarization direction perpendicular to the $[HHL]$ plane, and the sample was rotated in 0.5$^\circ$ steps over a total of 250$^\circ$. The data is subsequently folded into a single quadrant of the $[HHL]$ plane and further symmetrized. We have further discussed this symmetrization process in the supplemental material of Ref.~\cite{Gaudet2019}. For each data set, we reduce the data in a manner that avoids adding artefacts arising from the subtraction of strong nuclear Bragg peaks. Allowed nuclear Bragg peaks are located at $\mathbf{Q} = (1,1,1), (2,2,2), (1,1,3), (2,2,0), (0,0,4)$, and symmetrically equivalent locations. The intensity at each Bragg peak location is masked in performing the temperature subtraction, and we then show the intensity at these masked Bragg peak locations as the average intensity of the surrounding points in reciprocal space.  

\section*{Appendix F: NLC Fitting to Integrated-$S(\mathbf{Q},T)$}

\begin{figure*}
\linespread{1}
\par
    \begin{center}
        \includegraphics[width=7in]{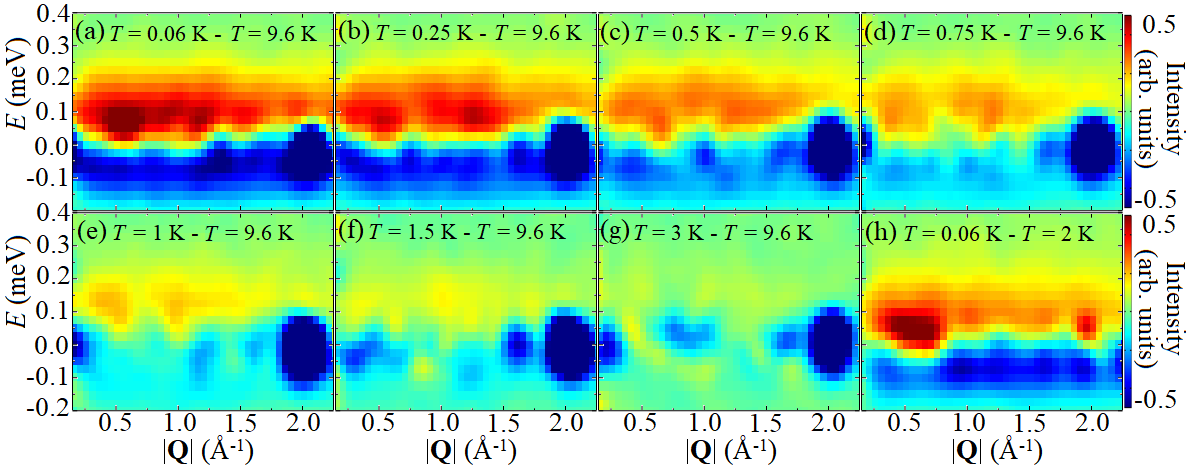}
        \par
        \caption{The temperature evolution of the low-energy inelastic neutron scattering from a powder sample of Ce$_2$Zr$_2$O$_7$. A data set measured at 9.6~K has been subtracted from a data set measured at $T$ =  0.06~K (a), 0.25~K (b), 0.5~K (c), 0.75~K (d), 1~K (e), 1.5~K (f), and 3~K (g). (h) The powder-averaged neutron scattering signal measured at $T$ =  0.06~K from a single-crystal sample of Ce$_2$Zr$_2$O$_7$, with a $T$ =  2~K data set subtracted, is shown for comparison.} 
        \end{center}
\label{Figure15}
\end{figure*}

The microscopic spin Hamiltonian parameters A and B can be employed in 3$^{\mathrm{rd}}$ order NLC calculations to calculate equal-time (energy-integrated) structure factors, and these can be compared to inelastic neutron scattering measurements on Ce$_2$Zr$_2$O$_7$.  The energy-integrated structure factor is:

\begin{equation}\tag{F1}
\begin{split}
    &S({\bf Q})= |f(|{\bf Q}|)|^2 \sum_{ij} \left( \hat{\bf z}_i  \cdot \hat{\bf z}_j - \frac{ (\hat{\bf z}_i  \cdot {\bf Q}) (\hat{\bf z}_j  \cdot {\bf Q})}{\lvert \bf Q \rvert^2} \right) \nonumber \\&\left(\cos^2(\theta) \langle S^{\tilde{x}}_i (-{\bf Q}) S^{\tilde{x}}_j ({\bf Q}) \rangle + \sin^2(\theta) \langle S^{\tilde{z}}_i (-{\bf Q}) S^{\tilde{z}}_j ({\bf Q}) \rangle \right) \nonumber , \\
\end{split}
\end{equation}

where $i,j,$ are sublattice indices and $f(|{\bf Q}|)$ is the magnetic form factor for Ce$^{3+}$. Averaging over directions at fixed magnitude $|{\bf Q}|= Q$, gives the powder structure factor $S(Q)$. We integrate over $Q = [0.46, 0.93]~\angstrom^{-1}$ and the result represents the energy-integrated neutron scattering response of a powder sample integrated over that momentum range, as a function of $T$. The structure factor at $T$ = 9.6~K is subtracted to replicate the background subtraction used in the experiment. Lines in Fig.~10 of the main text represent the powder integrated equal-time structure factor calculated in 3$^{\mathrm{rd}}$ order NLC using parameter sets A with $\theta=0$, and B with $\theta=0.561$~radians. 

We compare the NLC calculations to the experimentally measured neutron scattering response of a powder sample, integrated over the energy range $[-0.2, 0.4]$~meV. This integration range is chosen to enclose the low-lying excitations of the system while avoiding unnecessary contamination to the temperature-subtracted signal, which often becomes more prevalent at higher energies. To further reduce the effect of noise on the experimental data, we integrate in momentum-transfer over the range $\lvert\textbf{Q}\rvert = [0.46, 0.93]~\angstrom^{-1}$. This integration range is chosen to avoid nuclear Bragg peaks while still enclosing the dominant portion of the measured magnetic signal. We find that parameter sets from region A correctly predict the observed increase in scattering over the range $\lvert\textbf{Q}\rvert = [0.46, 0.93]~\angstrom^{-1}$ with decreasing temperature, while parameter sets from region B do not, as shown in Fig.~10 of the main text. 

Fig.~15(a) (15(b), 15(c), 15(d), 15(e), 15(f), 15(g)) shows the temperature-difference neutron scattering spectra measured from an annealed powder sample of Ce$_2$Zr$_2$O$_7$ for a $T$ = 0.06~K (0.25~K, 0.5~K, 0.75~K, 1~K, 1.5~K, 3~K) data-set with a $T$ = 9.6~K data-set used a background. The data in Fig.~15(a) (15(c), 15(g)) is also shown in Fig.~9(a) (9(b), 9(c)) with different $\lvert\textbf{Q}\rvert$-range and $\lvert\textbf{Q}\rvert$, $E$ pixel size. These data sets were used to compute the measured integrated-intensity over the energy range $[-0.2, 0.4]$~meV and the momentum range $[0.46,0.93]~\angstrom^{-1}$, forming the data points shown in Fig.~10 of the main text. 

\section*{Appendix G: Comparison of Temperature Dependence of $C_\mathrm{mag}$ with Powder-Averaged Inelastic Neutron Scattering}

\begin{figure}[t]
\linespread{1}
\par
\includegraphics[width=3.4in]{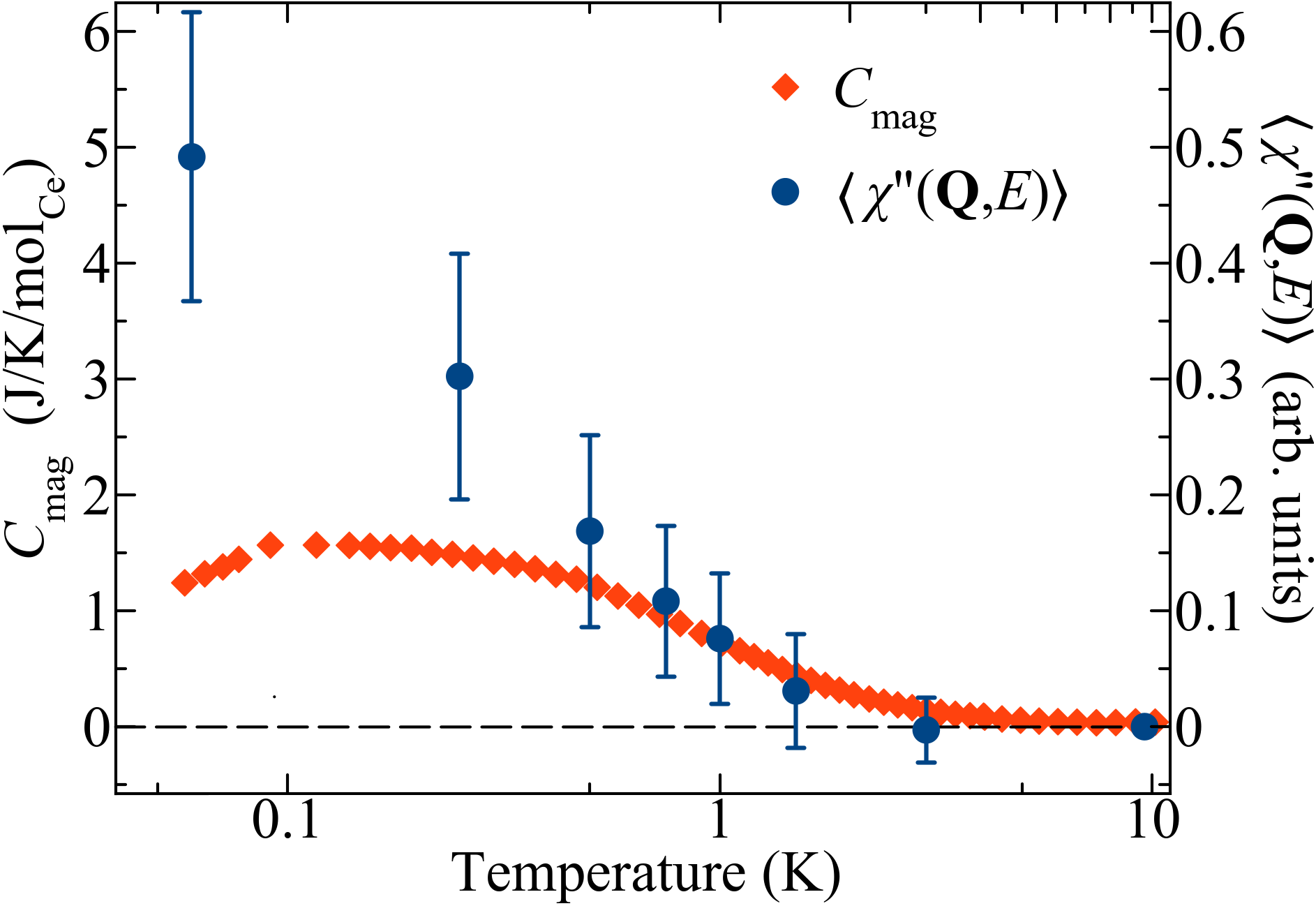}
\par
\caption{ The low energy dynamic susceptibility, $\chi^{\prime\prime}(\textbf{Q},E)$, averaged over $\lvert\textbf{Q}\rvert = [0.46, 0.93]~\angstrom^{-1}$ and $E = [0, 0.15]$~meV, is plotted alongside the measured $C_\mathrm{mag}$ for the Ce$_2$Zr$_2$O$_7$ sample of the present work.} 
\label{Figure16}
\end{figure}
 
The new measured $C_\mathrm{mag}$ data also allows for a comparison with the temperature-dependent inelastic neutron scattering signal measured on an annealed powder sample of Ce$_2$Zr$_2$O$_7$~\cite{Gaudet2019}. Specifically, we compare the $C_\mathrm{mag}$ data with the imaginary part of the dynamic spin susceptibility, $\chi^{\prime\prime}(\textbf{Q},E)$, calculated from our previously reported neutron data. As shown in Fig.~15(a-g), a signal with the approximate energy range $E = [0, 0.15]$~meV is seen to onset in the inelastic neutron scattering spectra with decreasing temperature. The dominant intensity within this signal was used to calculate $\langle\chi^{\prime\prime}(\textbf{Q},E)\rangle$ for each temperature, giving rise to the data points shown in Fig.~16 and allowing us to further examine the temperature dependence of the measured neutron scattering signal. $\chi^{\prime\prime}(\textbf{Q},E)$ was calculated via $\chi^{\prime\prime}(\textbf{Q},E) = S_0(\textbf{Q},E,T)(1-e^{-E/k_\mathrm{B}T})$, where $S_0(\textbf{Q},E,T)$ = $S(\textbf{Q},E,T)- S(\textbf{Q},E,$ $T$ = 9.6~K). This subtraction is used to isolate the magnetic contribution to the measured neutron scattering spectra, and assumes that $\chi^{\prime\prime}(\textbf{Q},E) = 0$ at $T$ = 9.6~K. This was used to calculate the average of $\chi^{\prime\prime}(\textbf{Q},E)$ over $\lvert\textbf{Q}\rvert = [0.46, 0.93]~\angstrom^{-1}$, $E = [0, 0.15]$~meV, denoted as $\langle\chi^{\prime\prime}(\textbf{Q},E)\rangle$. As shown in Fig.~16, the temperature onset of $\langle\chi^{\prime\prime}(\textbf{Q},E)\rangle$ coincides well with that of the broad-hump in $C_\mathrm{mag}$, and $\langle\chi^{\prime\prime}(\textbf{Q},E)\rangle$ continues to grow, separating from $C_\mathrm{mag}$, below $T$ $\sim$0.3~K. 
 
Recent theory work on the XYZ model Hamiltonian with $J_{\tilde{x}}$ = $J_{\tilde{y}}$ (which is a relevant approximation for the best-fitting exchange parameters found in this work) has predicted that a U(1) quantum spin ice ground state can be realized upon decreasing temperature through a classical spin ice regime~\cite{Kato2015,Huang2018a,Huang2018b}. Furthermore, these works predict that a broad hump in $C_\mathrm{mag}$ onsets slowly on entrance into the classical spin ice regime upon decreasing temperature. This prediction is consistent with the coincidence of the temperature onsets of $\langle\chi^{\prime\prime}(\textbf{Q},E)\rangle$ and $C_\mathrm{mag}$ shown in Fig.~16.

\begin{figure*}
\linespread{1}
\par
    \begin{center}
        \includegraphics[width=7in]{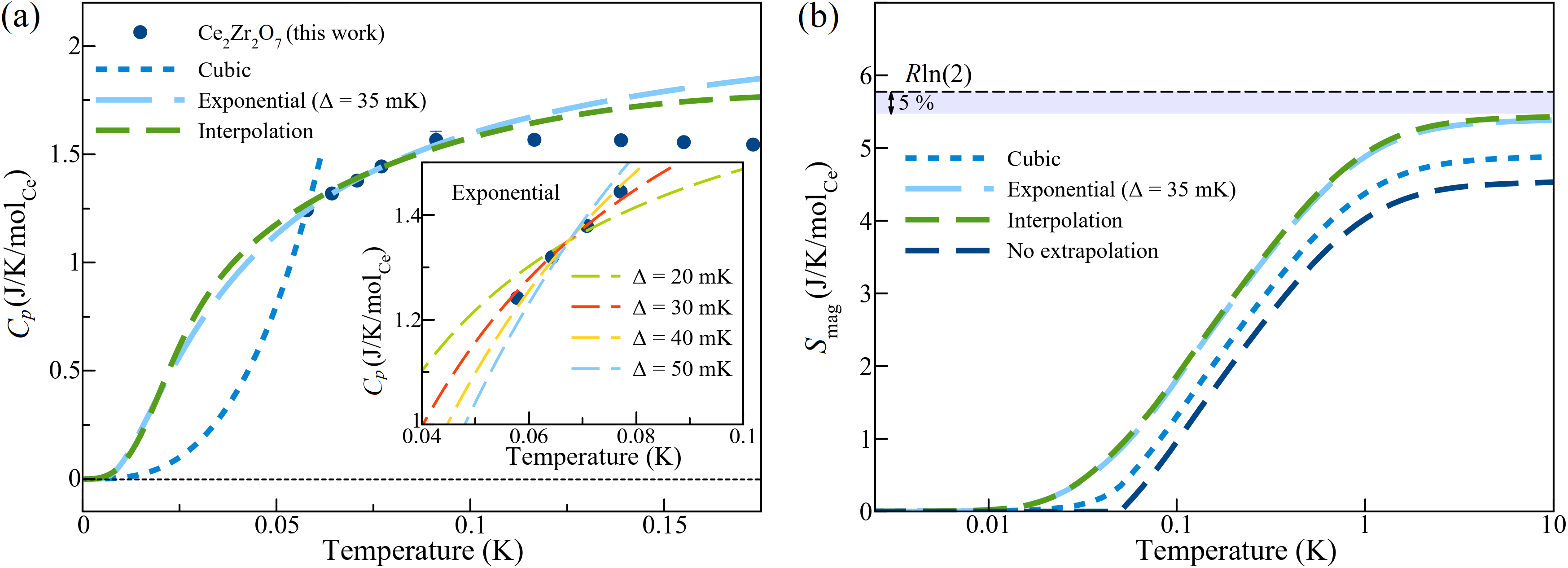}
        \par
        \caption{(a) The best-fitting naive cubic and exponential extrapolations to the measured $C_\mathrm{mag}$ data, as well as the low-temperature extrapolation which is consistent with the NLC fit for $T$ $>$ $\sim$0.5~K and a $T^3$ $C_\mathrm{mag}$ at sufficiently low temperature, as described in the text. A simple best-fitting cubic extrapolation forms a sharp cusp-like connection with the data while the simple best-fitting exponential extrapolation and the interpolation (as discussed in the text) both run smoothly through the lowest-temperature data points. The inset to (a) shows the simple exponential extrapolations for different values of the gap energy. Such an analysis yields an estimate of $\Delta$ = 35(5) mK for the gap energy. (b) A comparison of the entropy recovered via $S_\mathrm{mag} = \int_{0}^T\frac{C_\mathrm{mag}}{T}dT$ using the different low-temperature extrapolation schemes of $C_\mathrm{mag}$ that are shown in (a).} 
        \end{center}
\label{Figure17}
\end{figure*}

\section*{Appendix H: Semiclassical Molecular Dynamics Calculation of $S(\lvert\mathbf{Q}\rvert, E, T)$}

Here we discuss the semiclassical molecular dynamics calculations of $S(\lvert\mathbf{Q}\rvert, E, T)$ that lead to the calculated spectra shown in Fig.~9(d-i) of the main text. First, classical Monte Carlo simulations were performed using the best fit A exchange parameters, to obtain an ensemble of spin configurations sampled at temperature $T$. We then use these configurations as initial configurations and solve the semiclassical Landau-Lifshitz equation $\frac{d}{dt}\mathbf{S}_i$ = - $\mathbf{S}_i \times \mathbf{h}_i$, where $\mathbf{h}_i$ is the effective magnetic field on the spin $\mathbf{S}_i$. The dynamical structure factor is obtained as the time-space Fourier transform of the time-evolved magnetic moments, averaged over the ensemble of initial states. 

The molecular dynamics solution computes the classical dynamics. That is, it treats the spins as classical magnetic moments precessing in their local field. To compare this to the (quantum) experiment or a theoretical method such as linear spin wave theory, one has to re-scale the classical calculation. This is because the classical dynamical structure factor is symmetric with respect to neutron energy-transfer $E$, and it vanishes as $T$ approaches zero for all $E > 0$. Neither of these is the case for the dynamical structure factor of the quantum system. Another, more quantitative, way to think about this is via the fluctuation-dissipation theorem by comparing the version for classical and quantum systems~\cite{Kubo1966}. In particular, for a classical system we get $(\beta E) S_\mathrm{classical}(\textbf{Q}, E, T) = \chi^{\prime\prime}(\textbf{Q},E, T)$ while for the quantum system it reads $(1 - e^{-\beta E})S_\mathrm{quantum}(\textbf{Q}, E, T) = \chi^{\prime\prime}(\textbf{Q},E, T)$, where $\beta = 1/(k_\mathrm{B} T)$. It is then reasonable to equate the imaginary part of the susceptibility, $\chi^{\prime\prime}(\textbf{Q},E, T)$, as this quantity is real and symmetric for both the classical and the quantum system. Furthermore, as shown in Ref.~\cite{Zhang2019}, $\chi_\mathrm{quantum}^{\prime\prime}$ = $\chi_\mathrm{classical}^{\prime\prime}$ within linear spin wave theory. Using the quantum and classical fluctuation dissipation theorem for the respective sides then yields,

\begin{equation}\tag{H1}
S_\mathrm{quantum}(\textbf{Q}, E, T) = \frac{\beta E}{1-e^{-\beta E}}S_\mathrm{classical}(\textbf{Q}, E, T), 
\end{equation}
\\
which is what we use to estimate the dynamical structure factor of the (quantum) experiment using our classical simulation. The dynamical structure factor is then powder-averaged to obtain $S_\mathrm{quantum}(\lvert\textbf{Q}\rvert, E, T)$, and convolved with the experimental resolution. In Fig.~9(d-i) of the main text, we show the calculated powder-averaged dynamical structure factor at 0.06~K, 0.5~K, and 3~K, with the powder-averaged dynamical structure factor at $T$ = 9.6~K subtracted from the result.

Note that Eq.~(H1) accounts for detailed balance, $S_\mathrm{quantum}(\mathbf{Q}, -E, T)$ = $e^{-\beta E}S_\mathrm{quantum}(\mathbf{Q}, E, T)$, since $S_\mathrm{classical}(\mathbf{Q}, E, T) = S_\mathrm{classical}(\mathbf{Q}, -E, T)$. Zhang \emph{et al}. (Ref.~\cite{Zhang2019}) derive the conversion factor $\beta E$ by comparing the classical spin wave theory at finite temperature with the quantum spin wave theory at zero temperature. It is thus valid in the case $\beta E \gg 1$, which is well fulfilled in their case, but not applicable to a large part of our energy and temperature range. However, note that our factor $\frac{\beta E}{1 - e^{-\beta E}}$ reduces to $\beta E$ for $\beta E \gg 1$, so our calculation is entirely consistent with this argument. 

\section*{Appendix I: Heat Capacity Measurements and Low Temperature $C_\mathrm{mag}$-Extrapolations}

Heat capacity measurements were performed on our single crystal Ce$_2$Zr$_2$O$_7$ sample, along with a polycrystalline sample of La$_2$Zr$_2$O$_7$, which is used as a 4\textit{f}$^0$ analogue of Ce$_2$Zr$_2$O$_7$. Heat capacity measurements on a polished single crystal of Ce$_2$Zr$_2$O$_7$ (smooth-surfaced pressed powder pellet of La$_2$Zr$_2$O$_7$) were carried out on a Quantum Design PPMS down to $T$ = 0.058~K ($T$ = 2.5~K) using the conventional quasi-adiabatic thermal relaxation technique. The heat capacity of La$_2$Zr$_2$O$_7$ is very small at $\sim$2.5~K, and there was no need to pursue measurements at lower temperatures.

We provide further details on the analysis of $C_\mathrm{mag}$'s approach to zero at $T$ = 0~K. Fig.~17(a) shows the results of fitting simple cubic and exponential extrapolations to the measured $C_\mathrm{mag}$ data, as well as the low-temperature extrapolation to $C_\mathrm{mag}$ which is based upon an interpolation between the results of NLC calculations at $T$ $>$ $\sim$0.5~K and a $T^3$ low temperature form appropriate to emergent photons in a U(1) QSL. These extrapolations are also shown in Fig.~3 and Fig.~11(a) of the main text, respectively. We label the latter extrapolation as ``interpolation" in Fig.~17 and the following discussion. This interpolation method is introduced in Ref.~\cite{Bernu2001} and discussed for the current context below. 

The interpolation method first involves performing a high temperature expansion of the magnetic heat capacity, $C_\mathrm{mag}(T)$, corresponding to the XYZ Hamiltonian and the A set exchange parameters, and then turning this into an expansion for the entropy density as a function of energy density, $s(e)$, around $e=0$. If $C_\mathrm{mag}(T)$ $\propto$ $T^3$ at low temperature, then for $e$ close to the ground state energy density $e_0$, $s(e) \propto (e-e_0)^{3/4}$. A Pad{\'e} approximant is used to interpolate between those two limits, to obtain $s(e)$ over the region $e$ = $[e_0, 0]$, which can then be converted to $C_\mathrm{mag}(T)$ over the range $T$ = $[0, \infty]$. This approach requires an estimate of the ground state energy per site, $e_0$. We treat this estimate as an adjustable parameter and set $e_0 = -0.385J_a$ for best agreement with experiment, which is in a physically plausible range. 

The approach based on $s(e)$ is generally better behaved than performing the interpolation on $C_\mathrm{mag}(T)$ directly, and it obeys the physical constraints on the total energy and entropy $\int_0^{\infty} C_\mathrm{mag}(T) dT = -e_0$, $\int_0^{\infty} \frac{C_\mathrm{mag}(T)}{T} dT =$   $R\ln(2)$, respectively, by construction. The choice of Pad{\'e} approximant $P(m,n)$ is constrained to $m+n \leq k$, where $k$ is the maximum order obtained for the high temperature expansion of $C_\mathrm{mag}(T)$. In our case $k=13$, and we take the approximant $P(7,6)$, again guided by best agreement with experiment. The estimate of $e_0$ and the choice of $m,n$ are the only adjustable parameters in the comparison, with the exchange parameters equal to the set A parameters (see main text or Tab.~1). The comparison between theory and experiment is good, particularly for the entropy curve $S_\mathrm{mag}(T)$, when one considers that the experimental entropy is missing $\sim$5\% of the expected $R\ln(2)$, due to Ce$^{4+}$ substitution which is not incorporated in the interpolation calculation. This demonstrates that the observed $C_\mathrm{mag}(T)$ can be consistent with a smooth crossover to a $T^3$ form, even though we do not reach the $T^3$ regime with the present experimental data. 

As shown in Fig.~17(a), the cubic extrapolation cannot be made to connect smoothly to the data at the lowest temperature data points, while the best-fitting exponential extrapolation and the interpolation both meet the data in a smooth manner. The inset to Fig.~17(a) shows the best-fitting simple exponential extrapolation when locking the gap energy to the values $\Delta$ = 20, 30, 40, and 50~mK, and a gap of $\Delta$ = 35(5)~mK results from such a naive analysis.

Using each of these extrapolations for $C_\mathrm{mag}$ in order to describe the data below the lowest temperature data point, we calculate the entropy recovered via $S_\mathrm{mag} = \int_{0}^T\frac{C_\mathrm{mag}}{T}dT$ and show the results in Fig.~17(b). As shown in Fig.~17(b), the best-fitting cubic extrapolation grossly underestimates the $R\ln(2)$ entropy associated with the CEF ground state doublet, while the exponential extrapolation and the interpolation both saturate to $R\ln(2)$ within the $\sim$5\% tolerance associated with the sample oxidation. 

\section*{Appendix J: NLC calculations and Disorder Averaging}

We use the NLC method to calculate thermodynamic quantities throughout this work. The method is described in Refs.~\cite{Tang2013,Tang2015,Applegate2012} (for example). Extensive quantities per site $\frac{\langle \mathcal{O} \rangle}{N}$ are represented as sums over contributions from clusters $c$:
\begin{equation}\tag{J1}
\frac{1}{N}\langle \mathcal{O} \rangle = \sum_{c} M_c W_c,
\label{eq:NLCsum}
\end{equation}

where $M_c$ is the cluster multiplicity, defined as the number of times $c$ can be embedded in the lattice, per site $N$. $W_c$ is the cluster weight:
\begin{equation}\tag{J2}
W_c =\langle \mathcal{O} \rangle_c-\sum_{s \subset c} W_s,
\label{eq:NLCweight}
\end{equation}
where $\langle \mathcal{O} \rangle_c$ is the expectation value of the quantity $\mathcal{O}$ taken from exact diagonalization on cluster $c$ with open boundary conditions. The second term in Eq.~(J2) is a sum over the weights of all subclusters of $c$. The sum in Eq.~(J1) is arranged in order of increasing cluster size. At high temperatures, terms from larger clusters vanish faster with increasing temperature and the series converges in the same manner as high temperature expansion. At sufficiently high temperature, one can then justify truncating the sum at finite cluster size. 

We employ a series of clusters starting with a single site and then all further clusters are constructed from full tetrahedra. The $n^{\mathrm{th}}$ order of the expansion incorporates clusters of size up to $n$ tetrahedra. We denote the $n^{\mathrm{th}}$ order calculation as ``NLC$n$''. For the heat capacity we have performed calculations up to 4$^{\mathrm{th}}$ order (NLC4). For the A parameter set we additionally performed NLC calculations of $C_\mathrm{mag}$ up to 7$^{\mathrm{th}}$ order (see inset of Fig. 6). The methodology for these 7$^{\mathrm{th}}$ order calculations is described in Ref. \cite{Schafer2020}. For $S(Q)$ and $S(\mathbf{Q})$ we have performed calculations up to 3$^{\mathrm{rd}}$ order (NLC3). For the susceptibility we have performed calculations up to 2$^{\mathrm{nd}}$ order (NLC2). 

For example, to estimate $S(\mathbf{Q})$ using Eq.~(F1) and the NLC method, we define for each cluster $c$ entering the expansion, the extensive quantities:
\begin{equation}\tag{J3}
\begin{split}
    &C_c({\bf Q})= |f(|{\bf Q}|)|^2 \sum_{i, j \in c} \left( \hat{\bf z}_i  \cdot \hat{\bf z}_j - 
    \frac{ (\hat{\bf z}_i  \cdot {\bf Q}) (\hat{\bf z}_j  \cdot {\bf Q})}{\lvert \bf Q \rvert^2} \right) \nonumber \\&\left(\cos^2(\theta) \langle S^{\tilde{x}}_i (-{\bf Q}) S^{\tilde{x}}_j ({\bf Q}) \rangle + \sin^2(\theta) \langle S^{\tilde{z}}_i (-{\bf Q}) S^{\tilde{z}}_j ({\bf Q}) \rangle \right) \nonumber , \\
\end{split}
\end{equation}

The NLC estimate of $S({\bf Q})$ is then:
\begin{equation}\tag{J4}
S_\mathrm{NLC}({\bf Q}) = \sum_c M_c W_c({\bf Q})
\end{equation}
where in this case (3$^{\mathrm{rd}}$ order NLC) we truncate the sum at a maximum cluster size of three tetrahedra. $M_c$ are the cluster multiplicities and $W_c({\bf Q})$
are the cluster weights
\begin{equation}\tag{J5}
W_c({\bf Q}) = C_c({\bf Q}) - \sum_{s  \subset c}  W_s({\bf Q})
\end{equation}
where the sum on the RHS is over subclusters of $c$.
 
To improve convergence of the $C_\mathrm{mag}$ calculations, we have used Euler transformation to the 3$^{\mathrm{rd}}$ and 4$^{\mathrm{th}}$ orders~\cite{Applegate2012}. The Euler transformed results at 3$^{\mathrm{rd}}$ and 4$^{\mathrm{th}}$ order are:
\begin{equation}\tag{J6}
\langle \mathcal{O} \rangle_{\sf Euler 3}= \frac{1}{2} \langle \mathcal{O} \rangle_{\sf NLC2} + \frac{1}{2} \langle \mathcal{O} \rangle_{\sf NLC3}
\end{equation}
and
\begin{equation}\tag{J7}
\langle \mathcal{O} \rangle_{\sf Euler 4}= \frac{1}{4} \langle \mathcal{O} \rangle_{\sf NLC2} + \frac{1}{2} \langle \mathcal{O} \rangle_{\sf NLC3}+\frac{1}{4} \langle \mathcal{O} \rangle_{\sf NLC4} ,
\end{equation}
where $\langle \mathcal{O} \rangle_{\sf NLC n}$ is the estimate of $\langle \mathcal{O} \rangle$ up to $n^{\mathrm{th}}$ order in NLC. 

For the susceptibility calculations we included a population of 14\% vacancies in the calculation, with disorder averaging. The disorder average can be taken as order-by-order in NLC. Since vacancy disorder is binary, the disorder average can be done exactly~\cite{Tang2015}. We have also performed heat capacity calculations with 5\% vacancy disorder, as a point of comparison to the calculations with the clean model. The fits of these calculations to the experimental data produce very similar results to those found for the clean model, as shown in Fig.~13.

\bibliography{PRX_Version_Bibliography.bib}
\end{document}